\documentclass[11pt]{JHEP3} 

\usepackage{epsfig,multicol,bbm}

\newcommand{\non}{\nonumber}
\newcommand{\del}{\partial}
\newcommand{\bec}[1]{\mbox{\boldmath{${#1}$}}}
\newcommand{\Vac}[1]{\bigg\langle{#1}\bigg\rangle}
\newcommand{\til}[1]{\tilde{#1}}
\def\gtsim{\mathrel{\hbox{\raise0.2ex
\hbox{$>$}\kern-0.75em\raise-0.9ex\hbox{$\sim$}}}}
\def\ltsim{\mathrel{\hbox{\raise0.2ex
\hbox{$<$}\kern-0.75em\raise-0.9ex\hbox{$\sim$}}}}

\title{$CP$ violation in the secluded $U(1)'$-extended MSSM}

\author{Cheng-Wei Chiang$^{1,2}$ and Eibun Senaha$^1$\\
	$^1$Department of Physics and Center for Mathematics and Theoretical
        Physics,
	National Central University, Chungli, Taiwan 320, R.O.C.\\
	$^2$Institute of Physics, Academia Sinica, Taipei, Taiwan 115.\\
	E-mail: \email{chengwei@phy.ncu.edu.tw, senaha@ncu.edu.tw}}

      \abstract{We study the Higgs sector of the secluded $U(1)'$-extended MSSM
        (sMSSM) focusing on $CP$ violation.  Using the one-loop effective
        potential that includes contributions from quarks and squarks in the
        third generation, we search for the allowed region under theoretical and
        experimental constraints.  It is found that the possible region for the
        electroweak vacuum to exist is quite limited, depending on the
        parameters in the model.  The masses and couplings of the Higgs bosons
        are calculated with/without $CP$ violation.  Even at the tree level,
        $CP$ violation is possible by complex soft SUSY breaking masses.
        Similar to the CPX scenario in the MSSM, the scalar-pseudoscalar mixing
        enables the lightest Higgs boson mass to become smaller than the $Z$
        boson mass while the coupling with the $Z$ boson is sufficiently
        suppressed to avoid the LEP experimental constraints.  However, unlike
        the CPX scenario, large $\mu$ and $A$ are not required for the
        realization of large $CP$ violation.  The typical spectrum of the SUSY
        particles is thus different.  We also investigate the possible upper
        bound of the lightest Higgs boson in the case of spontaneous $CP$
        violation.  The maximal value of it can reach above 100 GeV with maximal
        $CP$-violating phases.}


\begin{document} 


%
%
\section{Introduction}

Many new physics models have been proposed to address the issue of the so-called
gauge hierarchy problem that cannot be resolved within the framework of the
standard model (SM).  Supersymmetric extensions of the SM have been paid much
attention as possible solutions to this problem.  In particular, the minimal
supersymmetric standard model (MSSM) can solve not only this problem but also
cosmological problems such as dark matter and baryon asymmetry of the Universe
and so on.  Nevertheless, the model still has an unattractive feature: the $\mu$
problem, where $\mu$ appears in the mass term of the higgsinos.  As long as no
special symmetry exist in the theory, the scale of $\mu$ is supposed to be the
grand unified theory (GUT)/Planck scale from the naturalness point of view.
However, once the electroweak symmetry is broken, the scale of $\mu$ should be
at about the $W$ boson mass.  One direction to provide a natural scale for $\mu$
is to introduce a gauge singlet field ($S$) into the MSSM.  Several variations
of this extension have been proposed: the next-to-MSSM (NMSSM)
~\cite{Ellis:1988er,NMSSM,Funakubo:2004ka}, the nearly MSSM
(nMSSM)~\cite{nMSSM,Balazs:2007pf}, the $U(1)'$-extended MSSM
(UMSSM)~\cite{UMSSM1,Cvetic:1997ky,UMSSM2}, and the secluded $U(1)'$-extended
MSSM (sMSSM)~\cite{Erler:2002pr,Han:2004yd}.  Comparisons among these
singlet-extended MSSM models can be found in Refs.~\cite{EMSSM}.  A common
feature in these models is that there is no fundamental $\mu$ term in the
superpotential.  After the symmetry breaking associated with the singlet field
$S$, the $\mu$ term is effectively generated by the product of the dimensionless
coupling and the vacuum expectation value (VEV) of $S$, and thus no fine tuning
is required.  Because of the introduction of singlet field(s), such models have
richer physics than the MSSM.

In this paper, we focus on the Higgs sector of the sMSSM with particular
emphasis on $CP$ violation.  The sMSSM is a string-inspired model whose particle
content of the Higgs sector comprises two Higgs doublets and four Higgs
singlets.  They are charged under the $SU(3)_C\times SU(2)_{L}\times
U(1)_{Y}\times U(1)'_{Q'}$ gauge symmetry.  Once the additional $U(1)$ symmetry
is introduced, a new gauge boson $Z'$ must exist in the model and can mix with
the ordinary $Z$ boson~\cite{Leike:1998wr,Langacker:2008yv}.  From the negative
results of $Z'$ search at LEP, the magnitude of the mixing angle between them
(denoted by $\alpha_{ZZ'}$) must be suppressed at $\mathcal{O}(10^{-3})$
level~\cite{Yao:2006px}.  The sMSSM provides an explanation for such a $Z$-$Z'$
hierarchy in a natural way.  If the $U(1)'$ symmetry is broken around the TeV
scale, the VEVs of the additional three Higgs singlets ($S_1, S_2, S_3$) are
expected to be of $\mathcal{O}$(TeV).  This makes $\alpha_{ZZ'}$ small enough to
escape from the current experimental bounds on the $Z'$ boson.

Due to the extension in the Higgs sector, it is possible to break the $CP$
symmetry explicitly and spontaneously at the tree level, which is forbidden in
the MSSM.  It is well known that the Kobayashi-Maskawa $CP$-violating
phase~\cite{Kobayashi:1973fv} in the SM is too small to generate sufficiently
large baryon asymmetry of the Universe as observed today~\cite{ewbg}.
Therefore, additional $CP$-violating phases are required for successful
baryogenesis.  So far, electroweak baryogenesis have been studied in the singlet
extended MSSM models: the NMSSM~\cite{EWBG_NMSSM}, the
nMSSM~\cite{Balazs:2007pf,EWBG_nMSSM}, the UMSSM~\cite{EWBG_UMSSM} and the
sMSSM~\cite{EWBG_sMSSM}.  A detailed analysis of the connection between $CP$
violation and baryogenesis, however, is beyond the scope of this paper.

In our analysis, we use the one-loop effective potential that includes
contributions from the third-generation quarks and squarks.  We search for the
parameter space allowed by imposing both theoretical and experimental
constraints on the model.  Owing to the presence of extra Higgs singlet fields,
the tadpole conditions defined by the first derivatives of the Higgs potential
do not always give the desired vacuum, $v=246$ GeV.  Therefore, we also
numerically check whether or not the minimum is located at 246 GeV.  We find
that the possible region for the electroweak vacuum is quite limited, depending
on the model parameters.

In the sMSSM, the only source of physical $CP$ violation at the tree level comes
from the relative phase between the soft SUSY breaking masses and the phases of
the Higgs fields.  We calculate the Higgs boson masses and the couplings between
the gauge bosons and Higgs bosons in the cases of explicit $CP$ violation (ECPV)
and spontaneous $CP$ violation (SCPV).  It is found that due to the new
$CP$-violating phases, the mass of the lightest Higgs boson can be smaller than
that of the $Z$ boson.  On the other hand, the coupling of the lightest Higgs
boson to the $Z$ boson is sufficiently suppressed, similar to the CPX scenario
in the MSSM~\cite{Pilaftsis:1999qt,Carena:2000yi,Carena:2002bb}.  Nonetheless,
the $\mu$ and $A$ parameters are not necessarily large in this model, making the
spectrum of SUSY particles different from the CPX scenario.

We also provide a bound on the lightest Higgs boson mass in the case of SCPV.
Depending on the mass of charged Higgs bosons, the upper bound can reach above
100 GeV with maximal $CP$ violation.

The paper is organized as follows.  In Section \ref{sec:model}, we introduce the
model and define the $CP$-violating phases in a reparametrization invariant way.
Theoretical and experimental constraints are studied in Section
\ref{sec:allowed}.  We examine the effects of $CP$ violation on the Higgs boson
masses and couplings in Section \ref{sec:CPV}.  In particular, the explicit
$CP$-violating case is presented in Subsection \ref{subsec:ECPV} and the
spontaneous $CP$-violating case in Subsection \ref{subsec:SCPV}.  The discussion
about electric dipole moments (EDMs) is presented in Subsection
\ref{subsec:EDM}.  Finally, we summarize the work in Section
\ref{sec:conclusion}.  Formulas of the Higgs boson masses are given in
Appendix~\ref{app:neu_MassMat}.

%
%
\section{The model}\label{sec:model}

\begin{table}[t]
\caption{Particle content in the Higgs sector of sMSSM}
\begin{center}
\begin{tabular}{|c|c|}
\hline
Higgs & $SU(3)_C\times SU(2)_{L}\times U(1)_{Y}\times U(1)'_{Q'}$ \\
\hline\hline
$H_d$ & $ \bigg(\bec{1}, \bec{2}, -1/2, Q_{H_d}\bigg)$ \\
$H_u$ & $ \bigg(\bec{1}, \bec{2},1/2, Q_{H_u}\bigg)$ \\
$S$ & (\bec{1}, \bec{1}, 0, $Q_S$) \\
$S_1$ & (\bec{1}, \bec{1}, 0, $Q_{S_1}$) \\
$S_2$ & (\bec{1}, \bec{1}, 0, $Q_{S_2}$) \\
$S_3$ & (\bec{1}, \bec{1}, 0, $Q_{S_3}$) \\ 
\hline
\end{tabular}
\end{center}
\label{tab:particle}
\end{table}

The particle content in the Higgs sector of sMSSM comprises two Higgs doublets
($H_d, H_u$) and four Higgs singlets ($S, S_1, S_2, S_3$)~\cite{Erler:2002pr}.
As listed in Table \ref{tab:particle}, each field is charged under the
$SU(3)_C\times SU(2)_{L}\times U(1)_{Y}\times U(1)'_{Q'}$ gauge symmetry.
Though it is desirable to have $U(1)'$ charges ($Q$'s) chosen to make the model
anomaly free, a complete analysis of anomaly cancellation is beyond the scope of
this paper \footnote{To be anomaly free, exotic chiral supermultiplets are
  generally required~\cite{Cvetic:1997ky,Erler:2000wu,Kang:2004bz}.  For our
  purpose, we assume that they are heavy enough not to affect the phenomenology
  at the electroweak scale.}.  Neither will we address the gauge coupling
unification issue here as it requires the knowledge of full particle spectrum in
the model.
Instead, we focus exclusively on the Higgs sector.  The model which we are
considering is extended so that no dimensionful parameter exists in the
superpotential $\mathcal{W}$:
\begin{eqnarray}
\label{eqn:superpo}
\mathcal{W}\ni-\epsilon_{ij}\lambda SH_d^iH_u^j-\lambda_SS_1S_2S_3 ~,
\end{eqnarray}
where $\lambda$ and $\lambda_S$ are the dimensionless couplings.  Unlike the
NMSSM, the $U(1)'$ symmetry forbids a cubic term in the superpotential which can
cause a domain wall problem if the $Z_3$ symmetry is broken spontaneously.  Once
the Higgs singlet $S$ develops a VEV, an effective $\mu$ term is generated by
$\mu_{\rm eff}=\lambda\langle S\rangle$.  Therefore, the scale of $\mu_{\rm
  eff}$ is determined by the soft SUSY breaking terms.  In
Eq.~(\ref{eqn:superpo}) only, there is no interaction between the secluded Higgs
singlet fields $S_{1,2,3}$ and the two Higgs doublets $H_{u,d}$ and singlet $S$.

The Higgs potential at the tree level is given by the $F$-, $D$- and soft SUSY
breaking terms:
\begin{eqnarray}
V_0=V_F+V_D+V_{\rm soft},
\end{eqnarray}
where each term reads
\begin{eqnarray}
V_F&=&|\lambda|^2\big\{|\epsilon_{ij}\Phi_d^i\Phi_u^j|^2+|S|^2(\Phi_d^\dagger\Phi_d
	+\Phi_u^\dagger\Phi_u)\big\}+|\lambda_S|^2(|S_1S_2|^2+|S_2S_3|^2+|S_3S_1|^2),\\
V_D&=&\frac{g_2^2+g_1^2}{8}(\Phi_d^\dagger\Phi_d-\Phi_u^\dagger\Phi_u)^2 
	+\frac{g_2^2}{2}|\Phi_d^\dagger\Phi_u|^2\non\\
	&&+\frac{g'^2_1}{2}\Big(Q_{H_d}\Phi_d^\dagger\Phi_d
	+Q_{H_u}\Phi_u^\dagger\Phi_u+Q_S|S|^2+\sum_{i=1}^3Q_{S_i}|S_i|^2\Big)^2,\\
V_{\rm soft}&=&m_1^2\Phi_d^\dagger\Phi_d+m_2^2\Phi_u^\dagger\Phi_u 
	+m_S^2|S|^2+\sum_{i=1}^3m_{S_i}^2|S_i|^2\non\\
	&&-(\epsilon_{ij}\lambda A_{\lambda}S\Phi_d^i\Phi_u^j 
	+\lambda_SA_{\lambda_S}S_1S_2S_3+m_{SS_1}^2SS_1+m_{SS_2}^2SS_2
	+m_{S_1S_2}^2S_1^\dagger S_2+{\rm h.c.}).\non\\
\end{eqnarray}
where $g_2$, $g_1$ and $g'_1$ are the $SU(2)$, $U(1)$ and $U(1)'$ gauge
couplings, respectively.  We will take $g'_1=\sqrt{5/3}g_1$ as motivated by the
gauge unification in the simple GUTs. The soft SUSY breaking masses $m_{SS_1}$
and $m_{SS_2}$ are introduced to break the two unwanted global $U(1)$
symmetries. This choice is called Model I, where
$Q_S=-Q_{S_1}=-Q_{S_2}=Q_{S_3}/2$ and $Q_{H_d}+Q_{H_u}+Q_S=0$.  Although the
other choice dubbed Model II is also possible, we will not pursue it in this
paper since there is no room for physical $CP$-violating phases in the
tree-level potential~\cite{Erler:2002pr}.  The secluded sector $(S_1, S_2, S_3)$
can interact with the ordinary ones $(H_d, H_u, S)$ through the $g'_1$ coupling,
$m_{SS_1}$ and $m_{SS_2}$.

In general, the following five parameters can be complex in the Higgs potential:
\begin{eqnarray}
\lambda A_\lambda,~\lambda_S A_{\lambda_S},~m^2_{SS_1},~m^2_{SS_2},~m^2_{S_1S_2}
\in \bec{C}.
\end{eqnarray}
After rephasing the Higgs fields, however, four of them can be made real and
only one $CP$-violating phase is physical.  In the following, we define the
$CP$-violating phase in a reparametrization invariant way.  It should be noted
that in the UMSSM no physical $CP$-violating phase can survive after rotating
the Higgs fields and, therefore, the $CP$ symmetry cannot be violated in the
tree-level Higgs potential.  We parameterize the Higgs fields as
\begin{eqnarray}
\Phi_d&=&
e^{i\theta_1}\left(
\begin{array}{c}
\frac{1}{\sqrt{2}}(v_d+h_d+ia_d) \\
\phi_d^-
\end{array}
\right),\quad 
\Phi_u=
e^{i\theta_2}\left(
\begin{array}{c}
\phi_u^+\\
\frac{1}{\sqrt{2}}(v_u+h_u+ia_u) 
\end{array}
\right), \\
S&=&\frac{e^{i\theta_S}}{\sqrt{2}}(v_S+h_S+ia_S), \quad
S_i=\frac{e^{i\theta_{S_i}}}{\sqrt{2}}(v_{S_i}+h_{S_i}+ia_{S_i}),\quad(i=1-3),
\end{eqnarray}
where $v=\sqrt{v^2_d+v^2_u}\simeq 246$ GeV. The nonzero $\theta$'s can break the
$CP$ symmetry spontaneously.  However, the $\theta$'s are not independent.  Here
we define the four gauge invariant phases by
\begin{eqnarray}
\varphi_1&=&\theta_S+\theta_{S_1},\quad \varphi_2=\theta_S+\theta_{S_2},\quad
\varphi_3=\theta_S+\theta_1+\theta_2,\quad
\varphi_4=\theta_{S_1}+\theta_{S_2}+\theta_{S_3}.
\label{SCPV-phases}
\end{eqnarray}
For later convenience, we also define $\varphi_{12}=-\varphi_1+\varphi_2$.
%
%
The first derivative of the Higgs potential with respect to each Higgs field
must vanish (tadpole conditions).  At the tree level, we obtain
\begin{eqnarray}
\frac{1}{v_d}\Vac{\frac{\del V_0}{\del h_d}}
&=&m_{1}^{2}+\frac{g_{2}^{2}+g_{1}^{2}}{8}(v_{d}^{2}-v_{u}^{2})-R_\lambda\frac{v_uv_S}{v_d}
	+\frac{|\lambda|^2}{2}(v_u^2+v_S^2)+\frac{g'^2_1}{2}Q_{H_d}\Delta=0,\non\\
\label{tad1}\\
\frac{1}{v_u}\Vac{\frac{\del V_0}{\del h_u}}
&=&m_{2}^{2}-\frac{g_{2}^{2}+g_{1}^{2}}{8}(v_{d}^{2}-v_{u}^{2})-R_\lambda\frac{v_dv_S}{v_u}
	+\frac{|\lambda|^2}{2}(v_d^2+v_S^2)+\frac{g'^2_1}{2}Q_{H_u}\Delta=0,\non\\
\label{tad2}\\
\frac{1}{v_S}\Vac{\frac{\del V_0}{\del h_S}}
&=&m_S^2-{\rm Re}(m_{SS_1}^2e^{i\varphi_1})\frac{v_{S_1}}{v_S}
	-{\rm Re}(m_{SS_2}^2e^{i\varphi_2})\frac{v_{S_2}}{v_S}-R_\lambda\frac{v_dv_u}{v_S} \non\\
&&+\frac{|\lambda|^2}{2}(v_d^2+v_u^2)+\frac{g'^2_1}{2}Q_{S}\Delta=0,\label{tad3}\\
\frac{1}{v_{S_1}}\Vac{\frac{\del V_0}{\del h_{S_1}}}
&=&m_{S_1}^2-{\rm Re}(m_{SS_1}^2e^{i\varphi_1})\frac{v_{S}}{v_{S_1}}
	-{\rm Re}(m_{S_1S_2}^2e^{i\varphi_{12}})\frac{v_{S_2}}{v_{S_1}}
	-R_{\lambda_S}\frac{v_{S_2}v_{S_3}}{v_{S_1}} \non\\
&&+\frac{|\lambda_S|^2}{2}(v_{S_2}^2+v_{S_3}^2)+\frac{g'^2_1}{2}Q_{S_1}\Delta=0,\label{tad4}\\
\frac{1}{v_{S_2}}\Vac{\frac{\del V_0}{\del h_{S_2}}}
&=&m_{S_2}^2-{\rm Re}(m_{SS_2}^2e^{i\varphi_2})\frac{v_{S}}{v_{S_2}}
	-{\rm Re}(m_{S_1S_2}^2e^{i\varphi_{12}})\frac{v_{S_1}}{v_{S_2}}
	-R_{\lambda_S}\frac{v_{S_1}v_{S_3}}{v_{S_2}} \non\\
&&+\frac{|\lambda_S|^2}{2}(v_{S_1}^2+v_{S_3}^2)+\frac{g'^2_1}{2}Q_{S_2}\Delta=0,\label{tad5}\\
\frac{1}{v_{S_3}}\Vac{\frac{\del V_0}{\del h_{S_3}}}
&=&m_{S_3}^2-R_{\lambda_S}\frac{v_{S_1}v_{S_2}}{v_{S_3}}
	+\frac{|\lambda_S|^2}{2}(v_{S_1}^2+v_{S_2}^2)
	+\frac{g'^2_1}{2}Q_{S_3}\Delta=0,\label{tad6}\\
\frac{1}{v_u}\Vac{\frac{\del V_0}{\del a_d}}
&=&\frac{1}{v_d}\Vac{\frac{\del V_0}{\del a_u}}=I_\lambda v_S=0,\label{tad7}\\
\Vac{\frac{\del V_0}{\del a_S}}
&=&{\rm Im}(m_{SS_1}^2e^{i\varphi_1})v_{S_1}+{\rm Im}(m_{SS_2}^2e^{i\varphi_2})v_{S_2}
	+I_\lambda v_dv_u=0,\label{tad8}\\
\Vac{\frac{\del V_0}{\del a_{S_1}}}
&=&{\rm Im}(m_{SS_1}^2e^{i\varphi_1})v_{S}-{\rm Im}(m_{S_1S_2}^2e^{i\varphi_{12}})v_{S_2}
	+I_{\lambda_S}v_{S_2}v_{S_3}=0,\label{tad9}\\
\Vac{\frac{\del V_0}{\del a_{S_2}}}
&=&{\rm Im}(m_{SS_2}^2e^{i\varphi_2})v_{S}+{\rm Im}(m_{S_1S_2}^2e^{i\varphi_{12}})v_{S_1}
	+I_{\lambda_S}v_{S_1}v_{S_3}=0,\label{tad10}\\
\Vac{\frac{\del V_0}{\del a_{S_3}}}&=&I_{\lambda_S}v_{S_1}v_{S_2}=0,\label{tad11}
\end{eqnarray}
with
\begin{eqnarray}
\Delta&=&Q_{H_d}v_d^2+Q_{H_u}v_u^2+Q_{S}v_S^2+\sum_{i=1}^3Q_{S_i}v_{S_i}^2,\\
R_\lambda&=&\frac{{\rm Re}(\lambda A_\lambda e^{i\varphi_3})}{\sqrt{2}},\qquad 
I_\lambda=\frac{{\rm Im}(\lambda A_\lambda e^{i\varphi_3})}{\sqrt{2}},\\
R_{\lambda_S}&=&\frac{{\rm Re}(\lambda_S A_{\lambda_S} e^{i\varphi_4})}{\sqrt{2}},\qquad 
I_{\lambda_S}=\frac{{\rm Im}(\lambda_S A_{\lambda_S} e^{i\varphi_4})}{\sqrt{2}},
\end{eqnarray}
where $\langle\cdots\rangle$ is defined such that all Higgs fluctuating fields
are taken to be zero.  Here all the Higgs VEVs are assumed to be nonzero.  For
some parameter sets, however, a global minimum can be located at the place where
some of the Higgs VEVs are zero. Of course, such a minimum cannot be found from
Eqs.~(\ref{tad1})-(\ref{tad11}).  We will discuss the method of minimum search
in Section~\ref{sec:allowed}. In the current investigation, we do not specify
any SUSY breaking scenario. Hence the soft SUSY breaking masses are given by the
tadpole conditions for the $CP$-even Higgs fields
Eqs.~(\ref{tad1})-(\ref{tad6}).  After solving the tadpole conditions for the
$CP$-odd Higgs fields from Eqs.~(\ref{tad7})-(\ref{tad11}), we find
\begin{eqnarray}
I_\lambda&=&I_{\lambda_S}=0,\label{cond_cpv1}\\
{\rm Im}(m^2_{SS_1}e^{i\varphi_1})
&=&{\rm Im}(m^2_{S_1S_2}e^{i\varphi_{12}})\frac{v_{S_2}}{v_S},\label{cond_cpv2}\\
{\rm Im}(m^2_{SS_2}e^{i\varphi_2})
&=&-{\rm Im}(m^2_{S_1S_2}e^{i\varphi_{12}})\frac{v_{S_1}}{v_S}\label{cond_cpv3}.
\end{eqnarray}
The $CP$-violating phases must satisfy Eqs.~(\ref{cond_cpv1})-(\ref{cond_cpv3})
for the vacuum.  As a convention, we choose the independent physical
$CP$-violating phase to be $\theta_{\rm phys}={\rm Arg}(m^2_{S_1S_2}) +
\varphi_{12}$.
%
%
\subsection{The mass matrix of the neutral Higgs bosons}

\begin{table}[t]
\caption{Physical Higgs bosons in the sMSSM}
\begin{center}
\begin{tabular}{|c|c|c|c|}
\hline
& $CP$-even Higgs bosons & $CP$-odd Higgs bosons 
& charged Higgs bosons\\
\hline\hline
CPC & $H_1,H_2,H_3,H_4,H_5,H_6$ & $A_1,A_2,A_3,A_4$ & $H^+, H^-$ \\
\hline
CPV & \multicolumn{2}{c|} {$H_1,H_2,H_3,H_4,H_5,H_6, H_7, H_8, H_9 ,H_{10}$} 
& $H^+, H^-$ \\
\hline
\end{tabular}
\end{center}
\label{tab:phys_particle}
\end{table}

The squared mass matrix of the neutral Higgs bosons is a $12 \times 12$
symmetric matrix taking the form
\begin{eqnarray}
\frac{1}{2}
\left(
\begin{array}{cc}
\bec{H}^T & \bec{A}^T
\end{array}
\right) {\cal M}_N^2
\left(
\begin{array}{c}
\bec{H} \\
\bec{A}
\end{array}
\right),\quad
{\cal M}_N^2=
\left(
\begin{array}{cc}
{\cal M}_S^2 & {\cal M}_{SP}^2 \\
({\cal M}_{SP}^2)^T & {\cal M}_P^2
\end{array}
\right) ~,
\label{neu_MassMat}
\end{eqnarray}
where
$\bec{H}^T\equiv(\bec{h}_O^T=(h_d~h_u~h_S)~\bec{h}_S^T=(h_{S_1}~h_{S_2}~h_{S_3}))$,
$\bec{A}^T\equiv(\bec{a}_O^T=(a_d~a_u~a_S)~\bec{a}_S^T=(a_{S_1}~a_{S_2}~a_{S_3}))$.
The subscripts $O$ and $S$ on $\bec{h}/\bec{a}$ denote `ordinary' and
`secluded', respectively.  In Table \ref{tab:phys_particle}, the physical Higgs
bosons in this model are listed for both the $CP$-conserving (CPC) and the
$CP$-violating (CPV) cases.  After the symmetry breaking, two neutral
Nambu-Goldstone bosons $G^0$ and $G'^0$ appear and are absorbed by the $Z$ and
$Z'$ bosons, respectively.  It is straightforward to decouple $G^0$ from the
squared mass matrix (\ref{neu_MassMat}) analytically by performing the rotation
\begin{eqnarray}
\left(
\begin{array}{c}
a_d \\
a_u
\end{array}
\right) = 
\left(
\begin{array}{cc}
\cos\beta & \sin\beta \\
-\sin\beta & \cos\beta
\end{array}
\right)
\left(
\begin{array}{c}
G^0 \\
a
\end{array}
\right),
\end{eqnarray}
where $\tan\beta \equiv v_u/v_d$.  We diagonalize the reduced $11 \times 11$
matrix $\tilde{\mathcal{M}}^2_N$ numerically: $O^T \tilde{\mathcal{M}}^2_NO
=\mbox{diag}(m^2_{G'^0},m^2_1,m^2_2,m^2_3,m^2_4,m^2_5,m^2_6,
m^2_7,m^2_8,m^2_9,m^2_{10})$, where $m_i < m_{i+1}~(i=1-9)$ and $O$ is an
orthogonal matrix.  The explicit expressions for the matrix elements in
Eq.~(\ref{neu_MassMat}) at the tree level are presented in
Appendix~\ref{app:neu_MassMat}.

A complex $m^2_{S_1S_2}$ and/or a nontrivial $\varphi_{12}$ can yield nonzero
mixing terms between $CP$-even and $CP$-odd Higgs bosons:
\begin{eqnarray}
\mathcal{M}^2_{SP} \propto {\rm Im}(m^2_{S_1S_2}e^{i\varphi_{12}}) ~.
\end{eqnarray}
This gives rise to broken $CP$ symmetry.
A detailed discussion about the $CP$-violating effects on the Higgs masses and
couplings will be presented in Subsections~\ref{subsec:ECPV} and
\ref{subsec:SCPV}.  In the $CP$-conserving case, $\mathcal{M}^2_{SP}=\bec{0}$
and Eq.~(\ref{neu_MassMat}) can be decomposed into two $6 \times 6$
sub-matrices.

%
%
Now we consider the one-loop corrections to the Higgs boson masses.  It suffices
for the current investigation to take into account the contributions of the
third-generation quarks ($t, b$) and squarks ($\tilde{t}_{1,2},
\tilde{b}_{1,2}$).
The one-loop effective potential is given by~\cite{MSSMHiggs_bound}
\begin{eqnarray}
V_1=\frac{N_C}{32\pi^2}\sum_{q=t,b}
\left[\sum_{a=1,2}\bar{m}^4_{\tilde{q}_a}\left(\ln\frac{\bar{m}^2_{\tilde{q}_a}}{M^2}
    -\frac{3}{2}\right)
  -2\bar{m}^4_q\left(\ln\frac{\bar{m}^2_q}{M^2}-\frac{3}{2}\right)\right] ~,
\end{eqnarray}
which is regularized using the $\overline{\rm DR}$-scheme.  Here $N_C$ denotes
the number of colors, $\bar{m}$'s are the background-field-dependent masses, and
$M$ is the renormalization scale.  We determine $M$ by the condition $\langle
V_1\rangle=0$, which implies
\begin{eqnarray}
\ln M^2 = 
\frac{\sum_{q}[\sum_{a}m^4_{\tilde{q_a}}\ln m^2_{\tilde{q_a}}-2m^4_q\ln m^2_q]}
	{\sum_{q}[\sum_{a}m^4_{\tilde{q_a}}-2m^4_q]}-\frac{3}{2} ~.
\end{eqnarray}
With the one-loop corrections, the tadpole conditions become
\begin{eqnarray}
0&=&\Vac{\frac{\partial V_0}{\partial \phi}}
+\frac{N_C}{16\pi^2}\sum_{q=t, b}\left[\sum_{a=1, 2}
  \bar{m}^2_{\tilde{q}_a}\Vac{\frac{\partial\bar{m}^2_{\tilde{q}_a}}{\partial \phi}}
  \left(\ln\frac{m^2_{\tilde{q}_a}}{M^2}-1\right)
  -2m_q^2\Vac{\frac{\partial \bar{m}^2_q}{\partial \phi}}
  \left(\ln\frac{m_q^2}{M^2}-1\right)\right] ~,\non\\
\end{eqnarray}
where $m^2=\langle\bar{m}^2\rangle$ and $\phi$ denotes all species of the Higgs
fields.  The one-loop corrections of the third-generation quarks and squarks to
the Higgs boson masses have exactly the same form as in the NMSSM.  The explicit
formulas can be found in Ref.~\cite{Funakubo:2004ka},
%
%
\subsection{The mass matrix of the charged Higgs bosons}
The charged Higgs sector is the same as in the MSSM.  Once the $\mu$ term in the
mass formula of the MSSM charged Higgs boson is replaced by the effective $\mu$
term, $\mu_{\rm eff}=\lambda v_Se^{i\theta_S}/\sqrt{2}$, we can readily obtain
the mass of the charged Higgs bosons in the sMSSM.  Its squared mass matrix is
given by
\begin{eqnarray}
\left(
\begin{array}{cc}
\phi_d^+ & \phi_u^+
\end{array}
\right) {\cal M}_\pm^2
\left(
\begin{array}{c}
\phi_d^- \\
\phi_u^-
\end{array}
\right) ~.
\label{ch_MassMat}
\end{eqnarray}
At the tree level, it follows from Eq.~(\ref{ch_MassMat}) that
\begin{eqnarray}
m_{H^\pm}^2
=\frac{1}{\sin\beta\cos\beta}\Vac{\frac{\del^2 V_0}{\del\phi_d^+\del\phi_u^-}}
=m_W^2+\frac{2R_\lambda}{\sin2\beta}v_S-\frac{|\lambda|^2}{2}v^2 ~.
\label{mch}
\end{eqnarray}
Due to the mixing terms between the Higgs doublets and singlets, the relation
between the charged Higgs boson mass and the $CP$-odd Higgs boson mass,
$m^2_{H^\pm}=m^2_W+m^2_A$ valid in the MSSM, breaks down in general.  In the
limit of $\lambda\to0$ and $v_S\to\infty$ with $\lambda v_S$ being fixed,
$m^{}_{SS_1}=m^{}_{SS_2}=0$ and without $CP$ violation, one of the $CP$-odd
Higgs boson masses is exactly given by $2R_\lambda v_S/\sin2\beta$.  The mass
relation in the MSSM is recovered in this particular case.

At the one-loop level, the mass formula of the charged Higgs bosons takes the
form~\cite{Carena:2000yi,Funakubo:2002yb}
\begin{eqnarray}
m^2_{H^\pm}&=&m^2_W+\frac{2R_\lambda v_S}{\sin2\beta}-\frac{|\lambda|^2}{2}v^2\non\\
&&+\frac{N_C}{16\pi^{2}\sin\beta\cos\beta}
\bigg[\bigg(\frac{h(m_{\til t_{1}}^{2})}{(m_{\til t_{1}}^{2}-m_{\til b_{1}}^{2})(m_{\til t_{1}}^{2}-m_{\til b_{2}}^{2})}+\frac{2m^2_tR_{t}v_S}{v^2\sin^2\beta}\bigg)f(m_{\tilde{t}_1}^2, m_{\tilde{t}_2}^2) \non\\
&&\hspace{3cm}+\bigg(\frac{h(m_{\til b_{1}}^{2})}{(m_{\til b_{1}}^{2}-m_{\til t_{1}}^{2})(m_{\til b_{1}}^{2}-m_{\til t_{2}}^{2})}+\frac{2m^2_bR_{b}v_S}{v^2\cos^2\beta}\bigg)f(m_{\tilde{b}_1}^2, m_{\tilde{b}_2}^2) \non\\
&&\hspace{3cm}-\frac{4m^2_{t}m^2_{b}}{v^2\sin\beta\cos\beta}f(m_t^2,
m_b^2)\bigg] ~,
\label{1loop_mch}
\end{eqnarray}
where $R_{t,b}={\rm Re}(\lambda A_{t,b}e^{i\varphi_3})/\sqrt{2}$, $A_{t,b}$ are
defined as the trilinear couplings in the soft SUSY breaking sector, and
$f(m^2_1,m^2_2)$ is defined by
\begin{eqnarray}
f(m_1^2, m_2^2) &=&
\frac{1}{m_1^{2}-m_2^{2}}
\left[m_1^2\left(\ln\frac{m_1^2}{M^2}-1\right)
  - m_2^2\left(\ln\frac{m_2^2}{M^2}-1\right)\right] ~.
\label{function-f}
\end{eqnarray}
The explicit form of $h(m^2)$ is given in Ref.~\cite{Funakubo:2002yb}.  As is
done in Ref.~\cite{Funakubo:2004ka}, $|A_\lambda|$ is determined by
Eq.~(\ref{1loop_mch}).  Therefore, we take $m_{H^\pm}$ as an input in our
analysis.

\section{Allowed region}\label{sec:allowed}
Finding an acceptable minimum of the Higgs potential is a nontrivial task even
at the tree level.  Even if we require the tadpole conditions and
positive-definiteness of the squared masses of the Higgs bosons, the global
minimum can be found at $v\neq 246$ GeV.  This is because of the presence of the
Higgs singlets in the Higgs potential.  In Ref.~\cite{Erler:2002pr}, the
following method is adopted to search for the electroweak vacuum.  First, the
soft SUSY breaking masses and the two trilinear $A$ terms ($A_\lambda$ and
$A_{\lambda_S}$) are taken at arbitrary values.  After finding a viable minimum,
all the given dimensionful parameters are rescaled so that $v=246$ GeV.  In this
method, all the Higgs VEVs are determined through the six tadpole conditions
(\ref{tad1})-(\ref{tad6}).  Therefore unlike the MSSM, $\tan\beta$ is an output.
Our method is equivalent to that, but the other way around.  Explicitly, we take
the Higgs VEVs as the inputs, and then perform the minimum search.  That is,
$v=246$ GeV is given in advance.  However, as we will see in what follows, the
desired electroweak vacuum does not always exist.  For some input parameters,
the location of $v=246$ GeV can be unstable and the true minimum would roll down
to another point that does not give $v = 246$ GeV.  Redefining such a minimum as
$v=246$ GeV by rescaling the Higgs VEVs is then inconsistent with the original
value of $\tan\beta$ that is scale independent.  Therefore, we discard such
cases and keep $\tan\beta$ as a fixed input.  Before showing the numerical
results of the minimum search, we consider theoretical and experimental
constraints in the following two subsections, respectively.

\subsection{Theoretical constraints}\label{subsec:th_const}
The effective potential at the tree level is
\begin{eqnarray}
\langle V_0 \rangle
&=&\frac{1}{2}m_1^2v_d^2+\frac{1}{2}m_2^2v_u^2+\frac{1}{2}m_S^2v_S^2
	+\sum_i\frac{1}{2}m_{S_i}^2v_{S_i}^2\non\\
&&-{\rm Re}(m_{SS_1}^2e^{i\varphi_1})v_Sv_{S_1}-{\rm Re}(m_{SS_2}^2e^{i\varphi_2})v_Sv_{S_2}
	-{\rm Re}(m_{S_1S_2}^2e^{i\varphi_{12}})v_{S_1}v_{S_2},\non\\
&&-R_\lambda v_dv_uv_S-R_{\lambda_S}v_{S_1}v_{S_2}v_{S_3}
	+\frac{g_2^2+g_1^2}{32}(v_d^2-v_u^2)^2 \non\\
&&+\frac{|\lambda|^2}{4}(v_d^2v_u^2+v_d^2v_S^2+v_u^2v_S^2)
	+\frac{|\lambda_S|^2}{4}(v_{S_1}^2v_{S_2}^2+v_{S_2}^2v_{S_3}^2+v_{S_3}^2v_{S_1}^2)
	+\frac{g'^2_1}{8}\Delta^2.	
\end{eqnarray}
In each direction of $v_S=v_{S_1}$ and $v_S=v_{S_2}$ with other VEVs being zero,
we demand the coefficients of the quadratic terms be positive so that the
effective potential is not unbounded from below:
\begin{eqnarray}
m_S^2+m_{S_{i}}^2-2{\rm Re}(m_{SS_i}^2e^{i\varphi_{i}})>0 ~, \quad i=1,2.
\end{eqnarray}
Next we consider the vacuum of the Higgs potential.  From the tadpole conditions
Eqs.~(\ref{tad1})-(\ref{tad11}), the vacuum of the tree-level potential takes
the form
\begin{eqnarray}
\langle V_0\rangle_{\rm vac}
&=&\frac{1}{2}R_\lambda v_dv_uv_S+\frac{1}{2}R_{\lambda_S}v_{S_1}v_{S_2}v_{S_3}
	-\frac{g_2^2+g_1^2}{32}(v_d^2-v_u^2)^2 \non\\
&&-\frac{|\lambda|^2}{4}(v_d^2v_u^2+v_d^2v_S^2+v_u^2v_S^2)
	-\frac{|\lambda_S|^2}{4}(v_{S_1}^2v_{S_2}^2+v_{S_2}^2v_{S_3}^2+v_{S_3}^2v_{S_1}^2)
	-\frac{g'^2_1}{8}\Delta^2 ~.
\label{V0_vac}
\end{eqnarray}
After eliminating $R_\lambda$ with Eq.~(\ref{mch}) and imposing $\langle
V_0\rangle_{\rm vac} < 0$, the upper bound on the charged Higgs boson mass is
obtained:
\begin{eqnarray}
m_{H^\pm}^2&<& m_W^2+\frac{2|\lambda|^2v_S^2}{\sin^22\beta}+m_Z^2\cot^22\beta
	-\frac{4R_{\lambda_S}}{v^2\sin^22\beta}v_{S_1}v_{S_2}v_{S_3} \non\\
	&&+\frac{2|\lambda_S|^2}{v^2\sin^22\beta}
	(v_{S_1}^2v_{S_2}^2+v_{S_2}^2v_{S_3}^2+v_{S_3}^2v_{S_1}^2)\label{up_mch}
	+\frac{g'^2_1}{v^2\sin^22\beta}\Delta^2\equiv(m^{\rm max}_{H^\pm})^2 ~.
\label{mch-max}
\end{eqnarray}

\begin{figure}[t]
\begin{center}
\includegraphics[width=6.5cm]{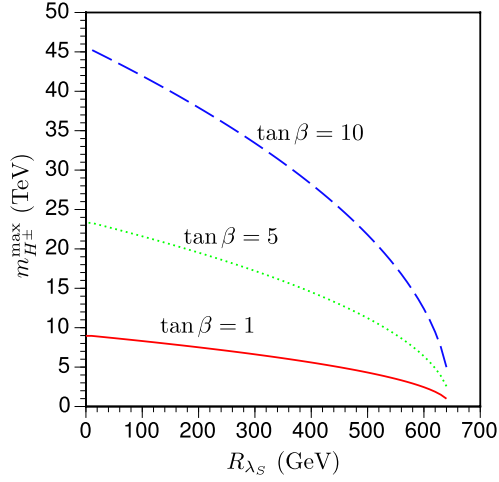}
\caption{The maximum of charged Higgs boson mass as a function of
  $R_{\lambda_S}$.  We take $v_S=300$ GeV, $v_{S_1}=v_{S_2}=v_{S_3}=3000$ GeV,
  and $\tan\beta=1$ (red solid line), 5 (green dotted line), 10 (blue dashed
  line).}
\label{fig:mch-max}
\end{center}
\end{figure}

As an example, we plot the maximal value of the charged Higgs boson mass as a
function of $R_{\lambda_S}$ in Fig.~\ref{fig:mch-max}.  We take $\lambda=-0.8$,
$\lambda_S=0.1$, $v_S=300$ GeV, $v_{S_1}=v_{S_2}=v_{S_3}=3000$ GeV, and
$\tan\beta=1$ (red solid line), 5 (green dotted line), 10 (blue dashed line).
The $CP$-violating phases are assumed to be zero.  Since the dominant terms are
proportional to $1/\sin^2 2\beta$ in $m^{\rm max}_{H^\pm}$, $\tan\beta=1$ gives
the smallest $m^{\rm max}_{H^\pm}$ for a fixed $R_{\lambda_S}$.  For
$R_{\lambda_S}>0$, the value of $m^{\rm max}_{H^\pm}$ decreases as
$R_{\lambda_S}$ increases.  We find a maximum of $R_{\lambda_S} \simeq 640$ GeV.
%
\subsection{Experimental constraints}\label{subsec:ex_const}
%
%
The $U(1)'$ charges of the Higgs fields can be constrained by the experimental
results of the $Z'$ boson search, namely, the lower bound on the $Z'$ boson mass
and the upper bound on the mixing angle between the $Z$ and $Z'$ bosons.  The
squared mass matrix of the $Z$ and $Z'$ bosons takes the form
\begin{eqnarray}
\mathcal{M}^2_{ZZ'}=
\left(
\begin{array}{cc}
m^2_Z & m_Zg'_1(Q_{H_d}\cos^2\beta-Q_{H_u}\sin^2\beta)v \\
m_Zg'_1(Q_{H_d}\cos^2\beta-Q_{H_u}\sin^2\beta)v & m^2_{Z'}
\end{array}
\right) ~,
\end{eqnarray}
where
\begin{eqnarray}
m^2_Z&=&\frac{g^2_2+g^2_1}{4}v^2,\\
m^2_{Z'}&=&g'^2_1\Big(Q_{H_d}^2v_d^2+Q_{H_u}^2v_u^2
	+Q_S^2v_S^2+\sum_i Q_{S_i}^2v_{S_i}^2\Big) ~.
\end{eqnarray}
The eigenvalues of the squared mass matrix and the mixing angle between the $Z$
and $Z'$ bosons are respectively given by
\begin{eqnarray}
m^2_{Z_{1,2}} &=&\frac{1}{2}\left[m^2_Z
	+m^2_{Z'}\mp\sqrt{(m^2_Z-m^2_{Z'})^2
	+g'^2_1(g^2_2+g^2_1)(Q_{H_d}v^2_d-Q_{H_u}v^2_u)^2} \right] ~, \\
\alpha^{}_{ZZ'}&=&\arctan\left(\frac{2m_Zg'_1(Q_{H_d}\cos^2\beta-Q_{H_u}\sin^2\beta)v}
	{m^2_{Z'}-m^2_Z}\right) ~.
\label{ZZp-mixing}
\end{eqnarray}

\begin{figure}[t]
\begin{center}
\includegraphics[width=6.5cm]{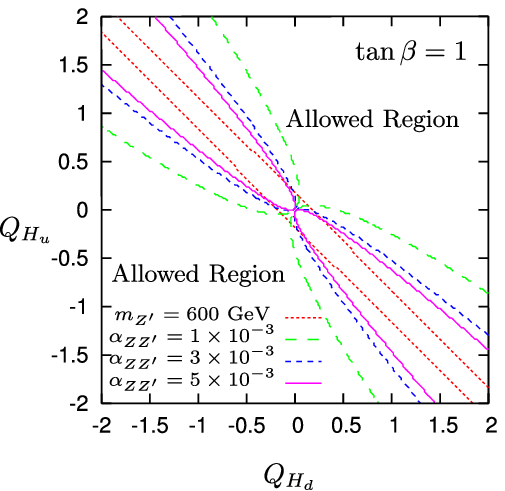}
\hspace{0.5cm}
\includegraphics[width=6.5cm]{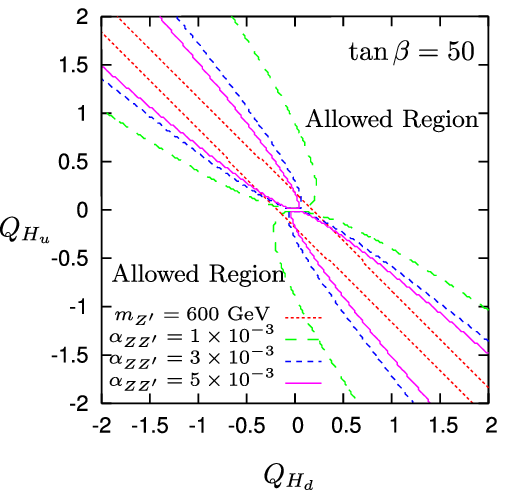}
\\
\vspace{0.5cm}
\includegraphics[width=6.5cm]{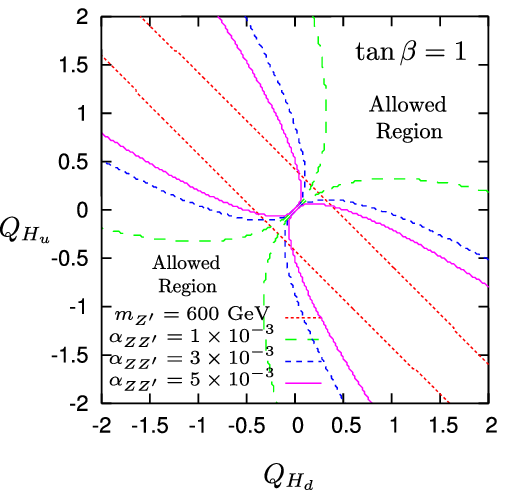}
\hspace{0.5cm}
\includegraphics[width=6.5cm]{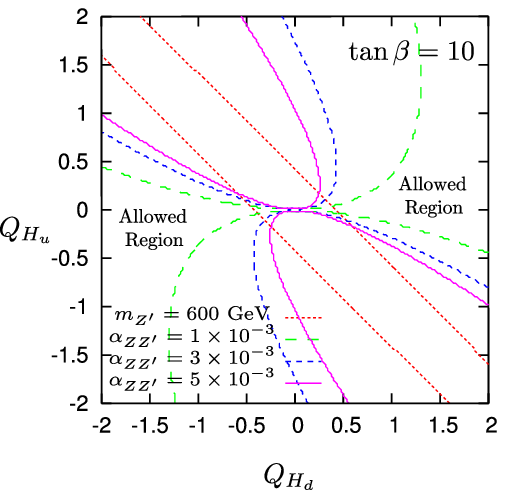}
\caption{Constraints from the lower bound on $m_{Z'}$ and the upper bound on
  $\alpha_{ZZ'}$ in the $Q_{H_u}$-$Q_{H_d}$ plane.  We take $v_S=300$ GeV,
  $v_{S_1}=v_{S_2}=v_{S_3}=3000$ GeV with $\tan\beta=1$ (upper left) and
  $\tan\beta=50$ (upper right), and $v_S=500$ GeV, $v_{S_1}=v_{S_3}=100$ GeV,
  $v_{S_2}=3000$ GeV with $\tan\beta=1$ (lower left) and $\tan\beta=10$ (lower
  right).}
\label{QHd-QHu}
\end{center}
\end{figure}

The experimental constraints on the $Z'$ boson are rather model-dependent.  Here
we adopt the typical bounds, $m_{Z'}>600$ GeV and $\alpha_{ZZ'} <
\mathcal{O}(10^{-3})$~\cite{Yao:2006px}.  In Figs.~\ref{QHd-QHu}, we plot the
$m_{Z'}=600$ GeV contour and curves for $\alpha_{ZZ'}=(1, 3, 5)\times 10^{-3}$
in the $Q_{H_u}$-$Q_{H_d}$ plane.  The other $U(1)'$ charges are determined by
the gauge invariance and the condition for breaking the two unwanted global
$U(1)$ symmetries as discussed above.  Here we show two examples: (A) $v_S=300$
GeV, $v_{S_1}=v_{S_2}=v_{S_3}=3000$ GeV with $\tan\beta=1$ (upper left figure)
and $\tan\beta=50$ (upper right figure); (B) $v_S=500$ GeV,
$v_{S_1}=v_{S_3}=100$ GeV, $v_{S_2}=3000$ GeV with $\tan\beta=1$ (lower left
figure) and $\tan\beta=10$ (lower right figure).  The red dotted lines give the
$m_{Z'}=600$ GeV contour, and the region in between represents $m_{Z'}\leq 600$
GeV.  The figures also show curves for $\alpha_{ZZ'}=1\times 10^{-3}$ (dashed
line in green), $\alpha_{ZZ'}=3\times 10^{-3}$ (dotted line in blue) and
$\alpha_{ZZ'}=5\times 10^{-3}$ (solid line in magenta).  In the region where
$Q_{H_d}$ and $Q_{H_u}$ have the same sign, the two terms in the off-diagonal
elements of $\mathcal{M}^2_{ZZ'}$ tend to cancel with each other.  The upper
right figures show that the $\tan\beta$ dependence on $Z'$ search constraints is
rather mild since the denominator in Eq.~(\ref{ZZp-mixing}) is relatively large
for case (A).  In the lower left figure, the covered areas of quadrants II and
IX have $\alpha_{ZZ'}>1\times10^{-3}$.  On the other hand, large portions of
quadrants I and III are not strongly constrained.  If we take $\tan\beta=10$,
the contours of $\alpha_{ZZ'}$ is distorted and the region around $Q_{H_d}\simeq
Q_{H_u}/\tan^2\beta$ becomes allowed.  In our numerical study, as long as one of
$v_{S_i}~(i=1-3)$ is taken to be at the TeV scale and $Q_{H_d}\simeq-Q_{H_u}$
does not hold, the constraints from the $Z'$ boson search can be easily avoided.
This supports the original motivation for the sMSSM as mentioned in the
Introduction.

%
%
According to the LEP experiments, the mass of the SM Higgs boson should be
larger than 114.4 GeV at 95 \% CL~\cite{Yao:2006px}.  However, this lower bound
cannot be directly applied to models beyond the SM due to the modification of
the Higgs coupling to the $Z$ boson ($g^{}_{HZZ}$).  When the Higgs boson masses
are smaller than 114.4 GeV, we require instead
\begin{eqnarray}
\xi^2 < k(m_{H_i}) ~,
\label{LEP-Higgs}
\end{eqnarray}
where $\xi = g^{}_{HZZ}/g^{\rm SM}_{HZZ}$ and $k$ is the 95 \% CL upper limit on
the $HZZ$ coupling and a function of the Higgs boson
mass~\cite{Barate:2003sz,Schael:2006cr}.  In our analysis, we do not consider
the processes $e^+e^- \to Z^* \to H_iH_j$.  They are expected to be less severe
in comparison with the processes $e^+e^- \to Z^* \to H_iZ$.

We also consider the $Z$ boson decays, $Z\to H_iH_j$ and $Z\to H_il^+l^-$ for
the light Higgs bosons, and require that:
\begin{eqnarray}
\sum_{i, j}\Gamma(Z\to H_iH_j)+\sum_i\Gamma(Z\to H_il^+l^-) < \Delta \Gamma_Z ~,
\end{eqnarray}
where $\Delta \Gamma_Z=2.0$ MeV is the 95 \% CL upper bound on the possible
additional decay width of the $Z$ boson~\cite{:2005ema}.

The other experimental constraints come from the lower bounds of the SUSY
particles.  The mass matrix of the charginos has the same form as in the MSSM if
we replace $\mu$ with $\mu_{\rm eff}$:
\begin{eqnarray}
\mathcal{M}_{\tilde{\chi}^\pm}=
\left(
\begin{array}{cc}
M_2 & -\sqrt{2}m_W\cos\beta \\
-\sqrt{2}m_W\sin\beta & \mu_{\rm eff}e^{i(\theta_1+\theta_2)}
\end{array}
\right),
\end{eqnarray}
where $M_2$ is the $SU(2)$ gaugino mass.  The physical $CP$-violating phase is
$\theta_{M_2}+\theta_\lambda+\varphi_3$, where $\theta_{M_2}$ and
$\theta_{\lambda}$ denote the arguments of $M_2$ and $\lambda$, respectively.
For the lower bound on the lightest chargino mass $\tilde{\chi}^\pm_1$, we
require $m_{\tilde{\chi}^\pm_1} > \sqrt{s}/2\simeq104$ GeV, where $\sqrt{s}$ is
the center-of-mass energy at LEP2~\cite{LEP2}.  On the other hand, the mass
bound on the neutralino, $m_{\tilde{\chi}^0}> 46$ GeV given in
Ref.~\cite{Yao:2006px} is rather model-dependent.  In fact, it is found that
$m_{\tilde{\chi}^0}\simeq 6$ GeV is allowed in the $R$-parity conserving MSSM
without gaugino mass unification~\cite{Bottino:2003iu}.  In the sMSSM, the
lightest neutralino can even be massless, almost a
singlino~\cite{neutralino_sMSSM}.  Therefore we will not put an explicit lower
bound on the mass of the lightest neutralino, and not require that the lightest
neutralino be a candidate for the cold dark matter of the Universe as well.

%
%
Now we consider extra contributions to the $\rho$ parameter.  It can be easily
shown that if a model has only Higgs doublets and singlets, $\rho=1$ at the tree
level.  As discussed before, as long as $\alpha_{ZZ'} < \mathcal{O}(10^{-3})$,
the deviation of the $\rho$ parameter from unity due to the $Z'$ boson is small
enough to evade the current experimental bound $\Delta\rho<2.0\times
10^{-3}$~\cite{Yao:2006px}.  Let us consider the one-loop corrections, focusing
particularly on the contributions of the physical Higgs bosons rather than
including all SUSY particles.  The correction to the $\rho$ parameter is given
by
\begin{eqnarray}
\Delta\rho=\frac{\Pi^T_{ZZ}(0)}{m^2_Z}-\frac{\Pi^T_{WW}(0)}{m^2_W} ~,
\end{eqnarray}
where $\Pi^T_{VV}(0)~(V=Z, W)$ are the transverse parts of the weak boson
self-energies at the zero momentum.  The Higgs boson contributions at the
one-loop level take the form
\begin{eqnarray}
\Delta\rho^{\rm Higgs}&=&\frac{G_F}{8\sqrt{2}\pi^2}
	\left[\sum_{i<j}g^2_{H_iH_jZ}B_5(m_{H_i}, m_{H_j})
	-\sum_i|g^{}_{H_iHW}|^2B_5(m_{H^\pm}, m_{H_i}) \right] ~,
\end{eqnarray}
with
\begin{eqnarray}
B_5(m_1, m_2)
&=&
\left\{
\begin{array}{l}
\displaystyle{ -\frac{1}{2}(m^2_1+m^2_2)+\frac{m^2_1m^2_2}{m^2_1-m^2_2}
\ln\frac{m^2_1}{m^2_2}
 	\quad(m_1\neq m_2)},\\
0 \quad(m_1=m_2) 
\end{array}
\right.,\\
g^{}_{H_iH_jZ}
&=& (O_{1i}O_{7j}-O_{1j}O_{7i})\sin\beta-(O_{2i}O_{7i}-O_{2j}O_{7i})\cos\beta ~, \\
g^{}_{H_iHW}
&=& O_{2i}\cos\beta-O_{1i}\sin\beta-iO_{7i},
\end{eqnarray}
where $G_F=1/(\sqrt{2}v^2)\simeq1.166\times10^{-5}$ (GeV$)^{-2}$.  Unlike the
MSSM, the custodial $SU(2)$ symmetry does not guarantee $\Delta\rho^{\rm
  Higgs}=0$ due to the contributions from the Higgs singlets.

Finally we comment in passing on the constraints from $B$ physics.  The
experimental results of $B_s\to \mu^+\mu^-$, $b\to s\gamma$ and $B^-_u \to
\tau^- \bar{\nu}_{\tau}$ can give a significant restriction on the parameter
space.  However, so long as we limit our interest to the low $\tan\beta$ region
($\ltsim 20$), constraints from the branching ratios of $B_s\to \mu^+\mu^-$ and
$B_u\to\tau\nu_{\tau}$ are less stringent.  The $b\to s\gamma$ process can be
important for the light charged Higgs bosons scenario, $m_{H^\pm}\ltsim 300$
GeV, in which case the contributions from the charged Higgs bosons and those of
the charginos have to cancel ~\cite{bsgam} in a way to be consistent with the
data~\cite{Barberio:2007cr}.  We leave the detailed analysis to another paper.

\subsection{Numerical evaluation}
\begin{figure}
\includegraphics[width=7cm]{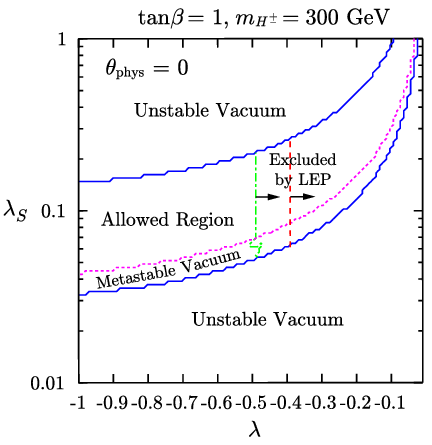}
\hspace{0.5cm}
\includegraphics[width=7cm]{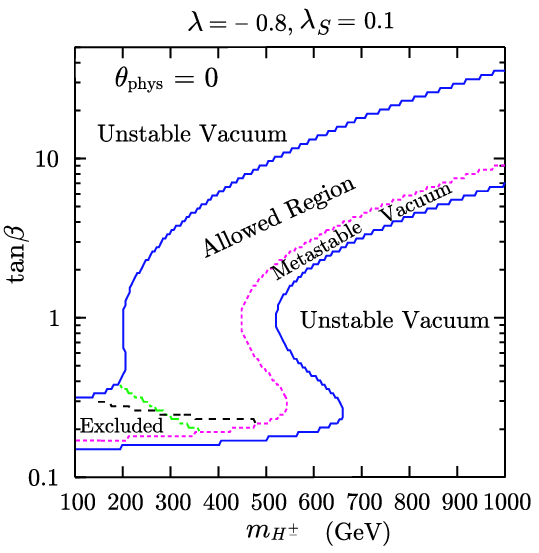}
\caption{The allowed region in the $\lambda_S$-$\lambda$ plane (left figure) and
  $\tan\beta$-$m_{H^\pm}$ plane (right figure).  We take $Q_{H_d}=Q_{H_u}=1$,
  $m^2_{SS_1}=m^2_{SS_2}=(500~{\rm GeV})^2$, $m^2_{S_1S_2}=-(50~{\rm GeV})^2$,
  $v_S=300$ GeV, $v_{S_1}=v_{S_2}=v_{S_3}=3000$ GeV.}
\label{bm1}
\end{figure}

Now we show the numerical results of the allowed regions in both case I and case
II.  We take
\begin{eqnarray}
&&Q_{H_d}=Q_{H_u}=1,\quad A_{\lambda_S}=A_{\lambda}(m_{H^\pm}),
\quad A_t=A_b=\mu_{\rm eff}/\tan\beta,\non\\
&&m_{\tilde{q}}=1000~{\rm GeV},\quad m_{\tilde{t}_R}=m_{\tilde{b}_R}=500~{\rm
  GeV},
\quad
M_2=200~{\rm GeV},
\end{eqnarray}
where $m_{\tilde{q}},~m_{\tilde{t}_R}$ and $m_{\tilde{b}_R}$ are the soft SUSY
breaking masses of squarks.  It should be noted that $A_{\lambda}$ is a function
of $m_{H^\pm}$, as given by Eq.~(\ref{1loop_mch}).  In Fig.~\ref{bm1}, the
allowed region is plotted in the $\lambda_S$-$\lambda$ plane (left figure) and
$\tan\beta$-$m_{H^\pm}$ plane (right figure).  The input parameters in Case I
are
\begin{eqnarray}
{\rm Case~I} :&&m^2_{SS_1}=m^2_{SS_2}=(500~{\rm GeV})^2,
~m^2_{S_1S_2}=-(50~{\rm GeV})^2,\non\\
&&v_S=300~{\rm GeV},~v_{S_1}=v_{S_2}=v_{S_3}=3000~{\rm GeV}.\label{caseI}
\end{eqnarray}
For the moment, all the $CP$-violating phases are assumed to be zero.  In the
left figure, we take $\tan\beta=1$ and $m_{H^\pm}=300$ GeV.  All the Higgs boson
masses are non-negative in the region between the two blue curves.  For fixed
$\lambda$, the depth of the vacuum decreases as $\lambda_S$ decreases and
eventually becomes higher than the origin, as can been seen from
Eq.~(\ref{V0_vac}).  The dotted curve in magenta corresponds to the critical
situation, below which the vacuum becomes metastable.  The region to the right
of the dotted-dashed line in green has been excluded by the condition
(\ref{LEP-Higgs}).  Likewise, the region to the right of the dashed line in red
is excluded by the chargino lower mass bound.  In the right figure, we take
$\lambda=-0.8,~\lambda_S=0.1$.  As in the left figure, $m^2_H\geq0$ is fulfilled
between the two blue curves, within which the vacuum becomes metastable below
the dotted curve in magenta. The region below the dotted-dashed curve in green
is excluded by the condition (\ref{LEP-Higgs}), and that below the dashed curve
in black by $\Delta\rho>2.0\times 10^{-3}$.  Since the Higgs singlets can affect
the lightest Higgs boson mass, the possibility $\tan\beta=1$ excluded in the
MSSM is experimentally allowed in our model.  On the contrary, the allowed
region is much more restricted by the conditions for the desired electroweak
vacuum.

\begin{figure}
\includegraphics[width=7cm]{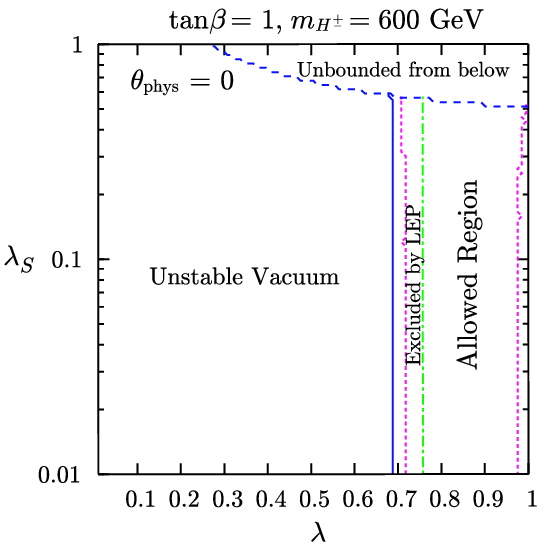}
\hspace{0.5cm}
\includegraphics[width=7cm]{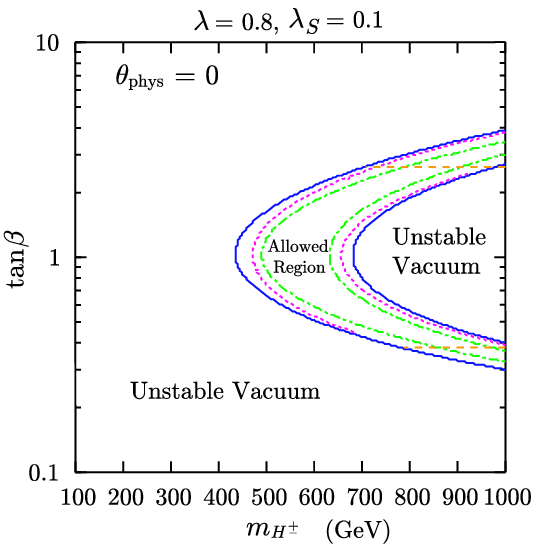}
\caption{The allowed region in the $\lambda_S$-$\lambda$ plane (left figure) and
  $\tan\beta$-$m_{H^\pm}$ plane (right figure).  We take $Q_{H_d}=Q_{H_u}=1$,
  $m^2_{SS_1}=(306~{\rm GeV})^2,~m^2_{SS_2}=(56~{\rm GeV})^2$,
  $m^2_{S_1S_2}=(100~{\rm GeV})^2$, $v_S=500$ GeV, $v_{S_1}=v_{S_2}=100$ GeV and
  $v_{S_3}=3000$ GeV.}
\label{bm2}
\end{figure}

In Fig.~\ref{bm2}, we consider
\begin{eqnarray}
{\rm Case~II} :&&m^2_{SS_1}=(306~{\rm GeV})^2,~m^2_{SS_2}=(56~{\rm GeV})^2,
~m^2_{S_1S_2}=(100~{\rm GeV})^2,\non\\
&&v_S=500~{\rm GeV},~v_{S_1}=v_{S_3}=100~{\rm GeV},~v_{S_2}=3000~{\rm GeV}.
\label{caseII}
\end{eqnarray}
In the left figure, we use $\tan\beta=1$ and $m_{H^\pm}=600$ GeV.  The region to
the left of the blue line is excluded by $m^2_{H}<0$, and that above the dashed
curve in blue results in the situation where $V=V_0+V_1$ is unbounded from
below.  In the region between the two lines in magenta, the vacuum is correctly
located at $v=246$ GeV.  However, the region to the left of the dotted-dashed
line in green is excluded by Eq.~(\ref{LEP-Higgs}).  The fact that $m_{H^\pm}$
in this case is larger than Case I implies that $R_\lambda$ is larger.  A small
$\lambda$ can make the vacuum metastable, as can be seen from
Eq.~(\ref{V0_vac}).  In the right figure, we take $\lambda=0.8$ and
$\lambda_S=0.1$.  The allowed region is inside the two dotted-dashed curves in
green and the two dashed lines in orange.  The dotted-dashed curves in green are
obtained from the critical value of the LEP bound (\ref{LEP-Higgs}) explained
above.  The dashed lines in orange correspond to $\alpha_{ZZ'} =
1\times10^{-3}$.  The parameter space is highly constrained in Case II.
%
\section{$CP$ violation}\label{sec:CPV}
In this section, we study the effects of $CP$ violation in the Higgs sector.  In
the MSSM, the $CP$-violating phase in the Higgs potential can be rotated away by
a field redefinition.  Hence there is no explicit $CP$ violation at the tree
level.  However, once the one-loop corrections from the squark sector to the
Higgs boson masses are taken into account, mixing terms between the $CP$-even
and $CP$-odd Higgs bosons are generated.  In a specific $CP$-violating case
called the CPX scenario, the effects of $CP$ violation is extremely enhanced,
and the Higgs phenomenology is drastically
changed~\cite{Pilaftsis:1999qt,Carena:2000yi,Carena:2002bb}.  The lightest Higgs
boson mass, for example, can become much smaller than the current LEP lower
bound due to the large $\mathcal{M}^2_{SP}$ in the squared mass matrix.  Its
coupling to the $Z$ boson, however, can be sufficiently suppressed to escape
from the LEP constraints~\cite{Schael:2006cr}.  Studies of ECPV have been done
in the NMSSM~\cite{Funakubo:2004ka,Matsuda:1995ta,Haba:1996bg},
nMSSM~\cite{Balazs:2007pf} and the UMSSM~\cite{Demir:2003ke,Ham:2007kc} as well.
Here we discuss both ECPV and SCPV in the sMSSM.

\subsection{Explicit $CP$ violation}\label{subsec:ECPV}
As discussed in Section~\ref{sec:model}, there is one $CP$-violating phase that
cannot be removed by rephasing the Higgs fields.  In fact, the nonzero
$CP$-violating phases are related to each other in the vacuum through the
tadpole conditions for the $CP$-odd Higgs fields.  At the one-loop level, we
find
\begin{eqnarray}
I_\lambda &=& -\frac{N_C}{8\pi^2v^2}
\left[\frac{m^2_t}{\sin^2\beta}f(m^2_{\tilde{t}_1},m^2_{\tilde{t}_2})
  + \frac{m^2_b}{\cos^2\beta}f(m^2_{\tilde{b}_1},m^2_{\tilde{b}_2})\right],
\label{cond_cpv1_loop} \\
I_{\lambda_S} &=& 0,
\label{cond_cpv2_loop} \\
{\rm Im}(m^2_{SS_1}e^{i\varphi_1})
&=&{\rm Im}(m^2_{S_1S_2}e^{i\varphi_{12}})\frac{v_{S_2}}{v_S},
\label{cond_cpv3_loop}\\
{\rm Im}(m^2_{SS_2}e^{i\varphi_2})
&=&-{\rm Im}(m^2_{S_1S_2}e^{i\varphi_{12}})\frac{v_{S_1}}{v_S}
\label{cond_cpv4_loop},
\end{eqnarray}
where $I_{t,b}={\rm Im}(\lambda A_{t,b}e^{i\varphi_3})/\sqrt{2}$.  If $I_t$ or
$I_b$ is nonzero, $I_\lambda$ can be nonzero as well at the one-loop level.
Nevertheless, we will focus exclusively on $CP$ violation peculiar to the sMSSM,
and take $I_t=I_b=0$ in what follows.  Since we have the relation
Eq.~(\ref{1loop_mch}), the sign of $R_\lambda$ is determined through
\begin{eqnarray}
{\rm sgn}(R_\lambda)={\rm sgn}\left(m^2_{H^\pm}-m^2_W
	+\frac{|\lambda|^2}{2}v^2-\Delta m^2_{H^\pm}\right),
\end{eqnarray}
where $\Delta m^2_{H^\pm}$ denotes the one-loop correction to the charged Higgs
boson mass.  On the contrary, there is a sign ambiguity in $R_{\lambda_S}$ at
this stage.  The positivity of the squared mass of the Higgs bosons gives us
$R_{\lambda_S}>0$ in most of the parameter space.  Now let us define
$\theta_{SS_1}={\rm Arg}(m^2_{SS_1}),~\theta_{SS_2}={\rm Arg}(m^2_{SS_2}),~
\theta_{S_1S_2}={\rm Arg}(m^2_{S_1S_2})$.  From Eqs.~(\ref{cond_cpv3_loop}) and
(\ref{cond_cpv4_loop}), it follows that
\begin{eqnarray}
\theta_{SS_1}&=&\sin^{-1}\left[\left|\frac{m^2_{S_1S_2}}{m^2_{SS_1}}\right|
	\frac{v_{S_2}}{v_S}\sin(\theta_{S_1S_2}+\varphi_{12})\right]-\varphi_1,
\label{thetaSS1}\\
\theta_{SS_2}&=&\sin^{-1}\left[-\left|\frac{m^2_{S_1S_2}}{m^2_{SS_2}}\right|
	\frac{v_{S_1}}{v_S}\sin(\theta_{S_1S_2}+\varphi_{12})\right]-\varphi_2.
\label{thetaSS2}
\end{eqnarray}
It should be noted that the arguments in the arcsines should be smaller than
one, imposing additional constraints on our input parameters.

The $CP$-violating phases show up in the mixing terms between $CP$-even and
$CP$-odd parts in the squared mass matrix (\ref{neu_MassMat}).  Let us
parameterize $\mathcal{M}^2_{SP}$ in terms of $3 \times 3$ block entries:
\begin{eqnarray}
\frac{1}{2}
\left(
\begin{array}{cc}
\bec{h}^T_O & \bec{h}^T_S
\end{array}
\right) {\cal M}_{SP}^2
\left(
\begin{array}{c}
\bec{a}_O \\
\bec{a}_S
\end{array}
\right),\quad
{\cal M}_{SP}^2=
\left(
\begin{array}{cc}
{\cal M}_{SP}^{(O)} & {\cal M}_{SP}^{(OS)} \\
\Big({\cal M}_{SP}^{(OS)}\Big)^T & {\cal M}_{SP}^{(S)}
\end{array}
\right).
\end{eqnarray}
After the conditions (\ref{cond_cpv3_loop}) and (\ref{cond_cpv4_loop}) are
applied, the entries are
\begin{eqnarray}
{\cal M}_{SP}^{(O)}&=&\bec{0}_{3\times3},
\quad {\cal M}_{SP}^{(OS)}={\rm Im}(m_{S_1S_2}^2e^{i\varphi_{12}})
\left(
\begin{array}{ccc}
0 & 0 & 0 \\
0 & 0 & 0 \\
\frac{v_{S_2}}{v_S} & -\frac{v_{S_1}}{v_S} & 0 \\
\end{array}
\right), \\
{\cal M}_{SP}^{(S)}&=&{\rm Im}(m_{S_1S_2}^2e^{i\varphi_{12}})
\left(
\begin{array}{ccc}
0 & 1 & 0 \\
-1 & 0 & 0 \\
0 &0 & 0
\end{array}
\right). 
\end{eqnarray}
If $\mathcal{M}^2_{SP}$ has a large portion in $\mathcal{M}^2_N$, the
$CP$-violating effects on the Higgs boson masses can be enhanced.  To achieve
this, we assume large values for ${\rm
  Im}(m^2_{S_1S_2}e^{i\varphi_{12}})v_{S_2}/v_S$ and ${\rm
  Im}(m^2_{S_1S_2}e^{i\varphi_{12}})v_{S_1}/v_S$ under the conditions
(\ref{thetaSS1}) and (\ref{thetaSS2}), rendering
\begin{eqnarray}
|m^2_{SS_1}|&\simeq&|m^2_{S_1S_2}|\frac{v_{S_2}}{v_S}, \\
|m^2_{SS_2}|&\simeq&|m^2_{S_1S_2}|\frac{v_{S_1}}{v_S},
\end{eqnarray}
for $\sin(\theta_{S_1S_2}+\varphi_{12})\simeq1$.  For the moment, we only
consider ECPV, and hence $\varphi_1=\varphi_2=0$.  We present two examples: one
being Case II as given in Eq.~(\ref{caseII}) and the other being Case III
specified by
\begin{eqnarray}
{\rm Case~III} :&&m^2_{SS_1}=(72~{\rm GeV})^2,~m^2_{SS_2}=(280~{\rm GeV})^2,
~m^2_{S_1S_2}=(100~{\rm GeV})^2,\non\\
&&v_S=300~{\rm GeV},~v_{S_1}=v_{S_3}=1500~{\rm GeV},~v_{S_2}=100~{\rm GeV}.
\label{caseIII}
\end{eqnarray}
We take $\tan\beta=1$ and $m_{H^\pm}=600$ GeV for Case II and $\tan\beta=1$ and
$m_{H^\pm}=300$ GeV for Case III.
\begin{figure}
\includegraphics[width=7cm]{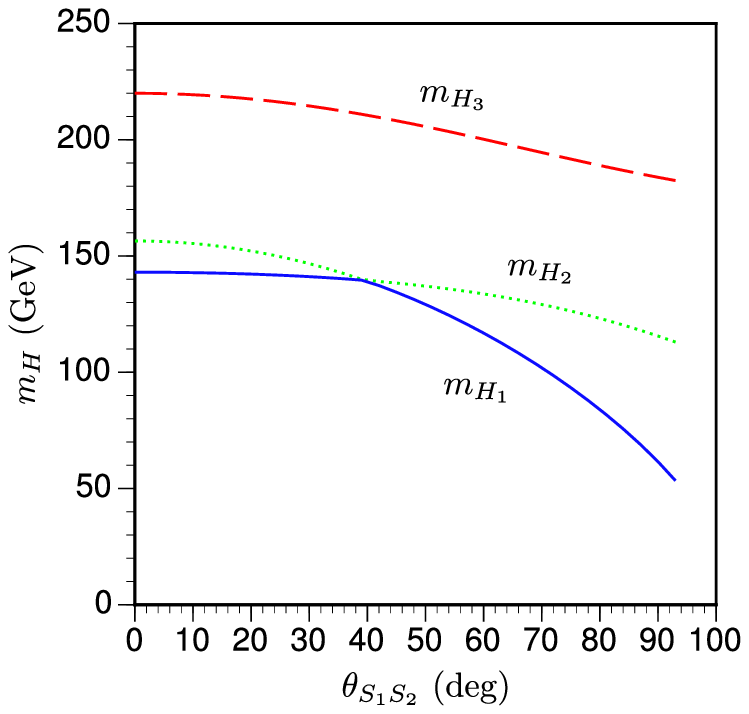}
\hspace{0.5cm}
\includegraphics[width=7cm]{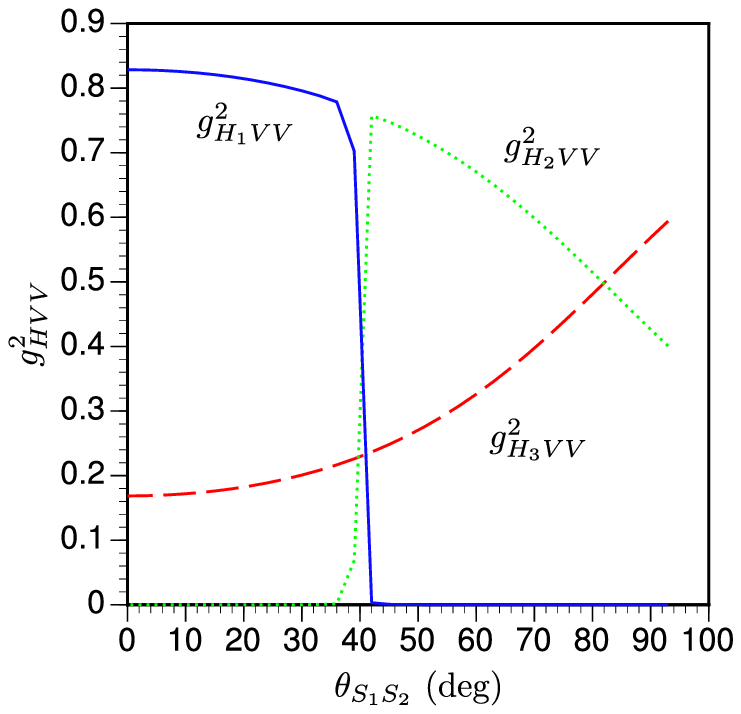}
\caption{The effects of the $CP$-violating phase on $m_H$ and $g^2_{HVV}$.  We
  take $m_{H^\pm}=600$ GeV, $\tan\beta=1$,~$|m^2_{SS_1}|=(306~{\rm GeV})^2$,
  $|m^2_{SS_2}|=(56~{\rm GeV})^2$, $|m^2_{S_1S_2}|=(100~{\rm GeV})^2$, $v_S=500$
  GeV, $v_{S_1}=v_{S_3}=100$ GeV, and $v_{S_2}=3000$ GeV.}
\label{LCPV1}
\end{figure}
In Fig.~\ref{LCPV1}, we plot $m_{H_i}$ and $g^2_{H_iVV}$~($i=1-3$) as functions
of $\theta_{S_1S_2}$ in Case II.  In the $CP$-conserving case,
$\theta_{S_1S_2}=0$, the second lightest Higgs boson is $CP$-odd because
$g^{}_{H_2VV}$ is zero.  Around $\theta_{S_1S_2} \simeq 40^\circ$, $H_1$ and
$H_2$ switch with each other and their $CP$ characters are exchanged, as can be
seen from the right figure in Fig.~\ref{LCPV1}.  As in the $CP$-violating MSSM,
due to the large off-diagonal terms $\mathcal{M}^2_{SP}$, $H_1$ can become
lighter than 114.4 GeV for $\theta_{S_1S_2}\gtsim 60^\circ$ with $g^2_{H_1VV}$
being highly suppressed.  This possibility cannot be excluded by the LEP
experimental results.  This does not seem to be typical in the $CP$-violating
NMSSM~\cite{Funakubo:2004ka}.  Although all the Higgs boson masses are positive
in the range $93^\circ$ $\ltsim\theta_{S_1S_2}\ltsim 102^\circ$, the vacuum is
metastable and is thus excluded.
\begin{figure}
\includegraphics[width=7cm]{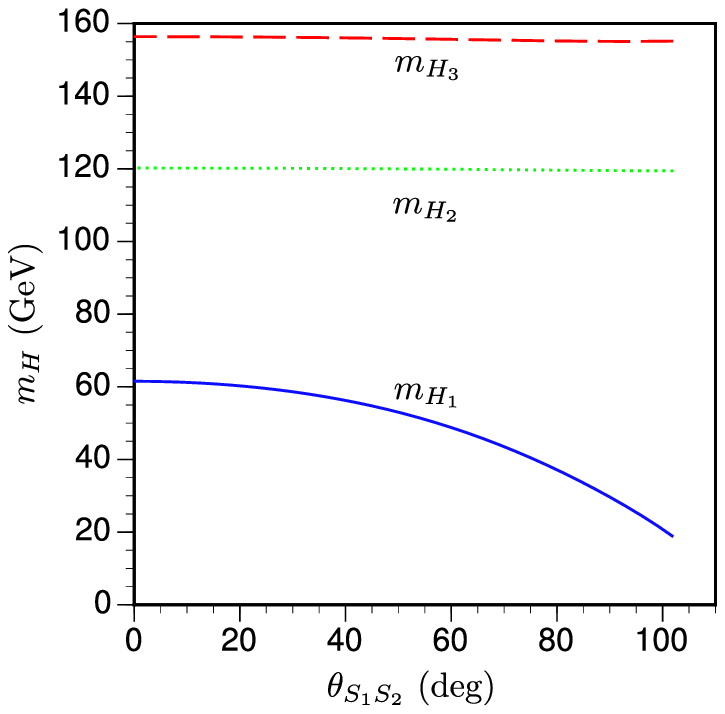}
\hspace{0.5cm}
\includegraphics[width=7.5cm]{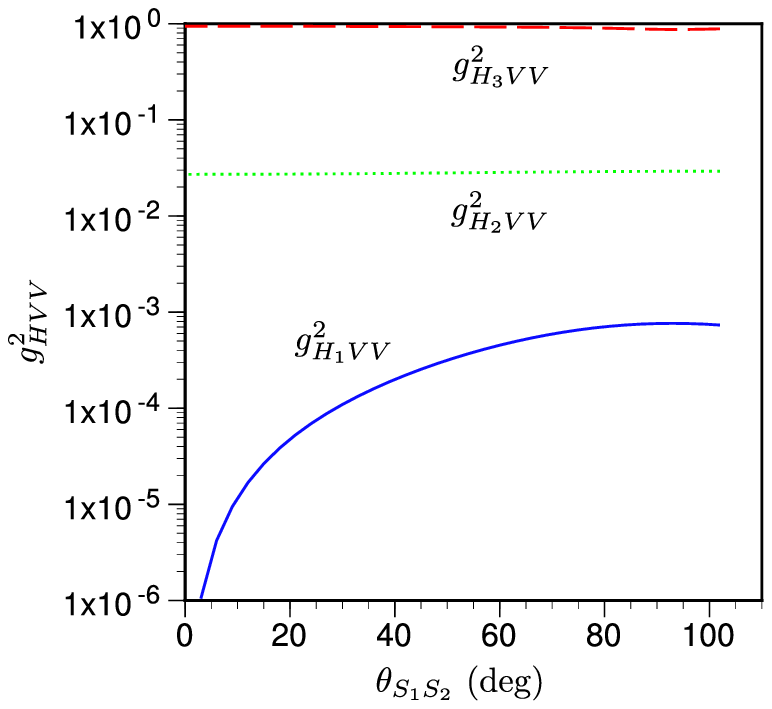}
\caption{The effects of the $CP$-violating phase on $m_H$ and $g^2_{HVV}$.  We
  take $m_{H^\pm}=300$ GeV, $\tan\beta=1$,~$|m^2_{SS_1}|=(72~{\rm GeV})^2$,
  $|m^2_{SS_2}|=(280~{\rm GeV})^2$, $|m^2_{S_1S_2}|=(100~{\rm GeV})^2$,
  $v_S=300$ GeV, $v_{S_1}=v_{S_3}=1500$ GeV, and $v_{S_2}=100$ GeV.}
\label{LCPV2}
\end{figure}
In Fig.~\ref{LCPV2}, we plot $m_{H_i}$ and $g^2_{H_iVV}~(i=1-3)$ as functions of
$\theta_{S_1S_2}$ for Case III.  When $\theta_{S_1S_2}=0$, $H_1$ is the $CP$-odd
Higgs boson since $g^{}_{H_1VV}=0$.  In this parameter set, $H_3$ is the SM-like
Higgs boson, corresponding to the decoupling limit in the MSSM.  Both $H_1$ and
$H_2$ are composed of almost singlet components.  The mass $m_{H_1}$ is always
smaller than the LEP bound when we vary $\theta_{S_1S_2}$, and can become as low
as 20 GeV around $\theta_{S_1S_2}=102^\circ$.  Since $g^2_{H_1VV}$ is less than
$10^{-3}$, the associated production cross section of $H_1$ with gauge bosons is
highly suppressed.  The masses and couplings of the other Higgs bosons are not
much affected by $CP$ violation.

\subsection{Spontaneous $CP$ violation}\label{subsec:SCPV}
In this subsection, we discuss the SCPV scenario.  If the model contains two
Higgs doublets, one of the Higgs VEVs can be complex in principle.  In the MSSM,
there is no room for the relative phase between the two Higgs doublets in the
potential in the SUSY limit due to $U(1)_{\rm PQ}$.  The only place where the
relative phase can show up is the quadratic mixing term between the two Higgs
doublets to break the SUSY softly.  After imposing the tadpole conditions, such
a phase disappears.  It is found that the one-loop corrections to the Higgs
potential can induce radiative SCPV~\cite{SCPV_MSSM}.  However, it leads to the
appearance of a light pseudoscalar ($m_A\ltsim 6$ GeV), which is already excluded
by the LEP experiments.  Many studies have already been done for SCPV in the
NMSSM with a $Z_3$ symmetry~\cite{Haba:1995aw,Romao:1986jy,
  Babu:1993qm,Ham:1999zs}.  According to Rom\~ao's No-Go
theorem~\cite{Romao:1986jy}, with certain radiative corrections in the Higgs
sector the condition for SCPV leads to a negative squared-mass mode in the Higgs
spectrum.  However, it is pointed out by Babu and Barr~\cite{Babu:1993qm} that
the large radiative corrections from the top/stop loops have not been taken into
account in the proof of the No-Go theorem.  The original saddle point in the
Higgs potential can become a minimum in this case and, therefore, the tachyonic
mode no longer appears.  In Ref.~\cite{Ham:1999zs}, the upper bound on the
lightest Higgs boson mass is found to be about 140 GeV in the case of SCPV where
the full one-loop corrections of top/stop have been included in their
calculations.  In the NMSSM without a $Z_3$ symmetry, the No-Go theorem cannot
be applied any more.  Hence, the SCPV scenario is viable even at the tree
level~\cite{SCPVwoZ3}.

\begin{figure}
\centerline{\includegraphics[width=5cm]{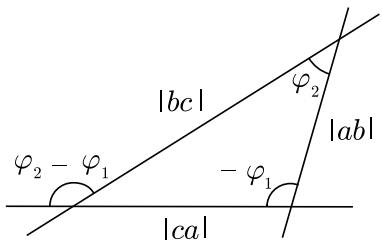}}
\caption{The representative solution for non-zero $\varphi_1$ and $\varphi_2$.}
\label{scpv_tri}
\end{figure}

In the sMSSM, SCPV is induced by the nonzero $\theta$'s that appear in the
quadratic terms of the Higgs potential.  This is also free from the No-Go
theorem.  To simplify our study, we assume that $m_{SS_1}^2$, $m_{SS_2}^2$,
$m_{S_1S_2}^2$, $\lambda A_\lambda$ and $\lambda_{S}A_{\lambda_S}$ are all real.
From the tadpole conditions (\ref{cond_cpv1_loop})-(\ref{cond_cpv4_loop}), we
find
\begin{eqnarray}
a\sin\varphi_1+b\sin\varphi_2&=&0,\label{cond_scpv1}\\
a\cos\varphi_1+b\cos\varphi_2&=&-\frac{ab}{c},\label{cond_scpv2}\\
\varphi_3=\varphi_4&=&0,
\end{eqnarray}
where $a=m_{SS_1}^2v_Sv_{S_1}$, $b=m_{SS_2}^2v_Sv_{S_2}$, and
$c=m_{S_1S_2}^2v_{S_1}v_{S_2}$.  When Eqs.~(\ref{cond_scpv1}) and
(\ref{cond_scpv2}) have solutions, they form a triangle as depicted in
Fig.~\ref{scpv_tri}.  The analytic solutions can be easily obtained:
\begin{eqnarray}
\cos\varphi_1&=&
\frac{1}{2}\left(\frac{bc}{a^2}-\frac{c}{b}-\frac{b}{c}\right),\\
\cos\varphi_2&=&
\frac{1}{2}\left(\frac{ac}{b^2}-\frac{a}{c}-\frac{c}{a}\right),\\
\cos(\varphi_1-\varphi_2)&=&
\frac{1}{2}\left(\frac{ab}{c^2}-\frac{b}{a}-\frac{a}{b}\right),
\end{eqnarray}
which give the $CP$-violating extremum.  The Higgs potential has the
$CP$-violating minimum when $ac/b<0$.

We can set $\theta_1=\theta_{S_3}=0$ without loss of generality in
Eq.~(\ref{SCPV-phases}).
Since $\varphi_3=\varphi_4=0$, it follows that
\begin{eqnarray}
\theta_2&=&-\frac{1}{2}(\varphi_1+\varphi_2),\quad 
\theta_S=\frac{1}{2}(\varphi_1+\varphi_2),\\
\theta_{S_1}&=&\frac{1}{2}(\varphi_1-\varphi_2),\quad 
\theta_{S_2}=-\frac{1}{2}(\varphi_1-\varphi_2).
\end{eqnarray}
\begin{figure}[t]
\includegraphics[width=7cm]{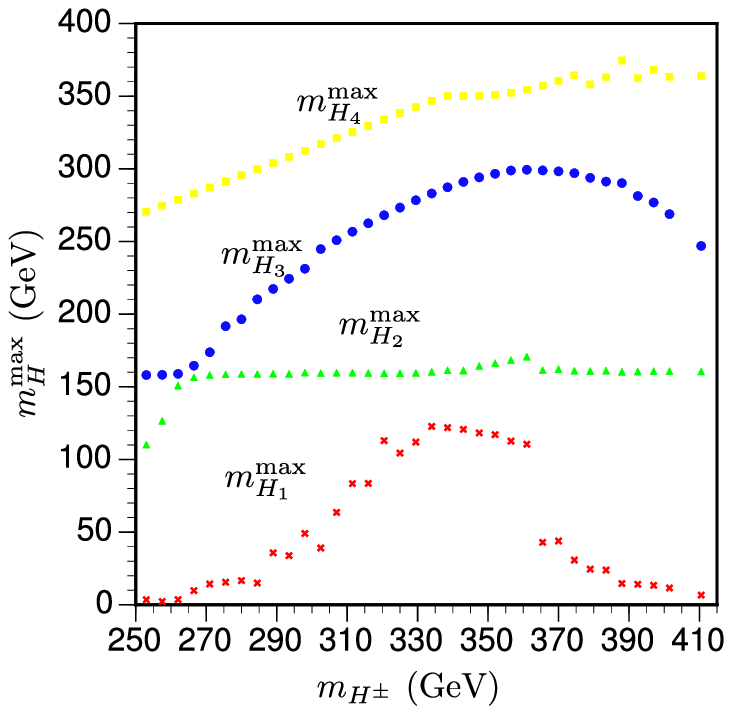}
\hspace{0.5cm}
\includegraphics[width=7cm]{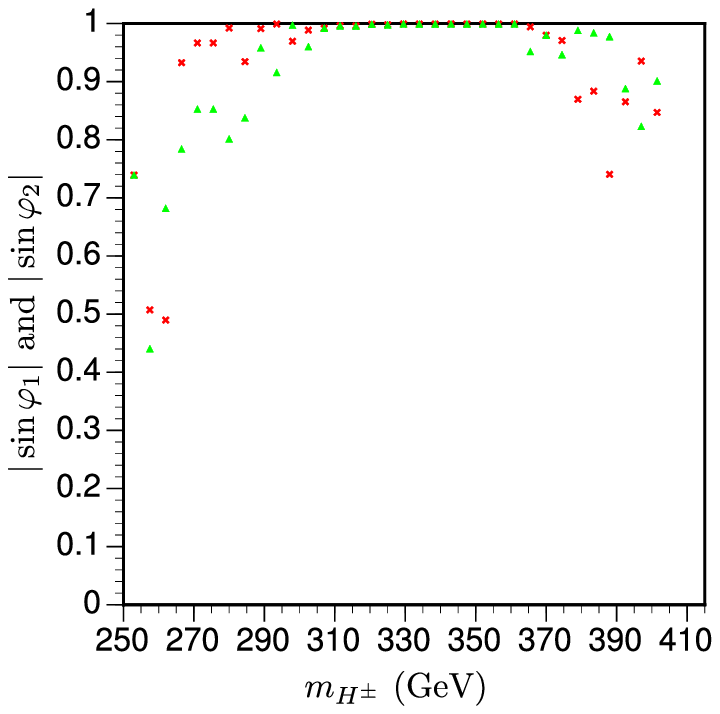}
\caption{The left plot shows the upper bounds on the four light neutral Higgs
  boson masses, $m^{\rm max}_{H_1}$ (cross in red), $m^{\rm max}_{H_2}$
  (triangle in green), $m^{\rm max}_{H_1}$ (circle in blue) and $m^{\rm
    max}_{H_1}$ (square in yellow), as functions of $m_{H^\pm}$.  The right plot
  shows $|\sin\varphi_1|$ and $|\sin\varphi_2|$.  The crosses in red are for
  $|\sin\varphi_1|$, and the triangles in green for $|\sin\varphi_2|$.}
\label{scpv_mHmax_bm1}
\end{figure}
We examine the possible maximal value of $m_{H}$ in the case of SCPV.  Since the
numerical minimum search is rather time-consuming, we do not conduct a complete
parameter scan.  Instead, we restrict ourselves to scan only the three soft SUSY
breaking masses in the following ranges:
\begin{eqnarray}
m^2_{SS_1}&=&m^2_{SS_2}=(10~{\rm GeV})^2-(1000~{\rm GeV})^2,\non\\
-m^2_{S_1S_2}&=&(1000~{\rm GeV})^2-(10~{\rm GeV})^2,
\end{eqnarray}
for fixed values of $m_{H^\pm}$.  The remaining parameters are chosen as
$\lambda=-0.8$, $\lambda_S=0.1$, $\tan\beta=1$, $v^{}_S=300~{\rm GeV}$, and
$v^{}_{S_1}=v^{}_{S_2}=v^{}_{S_3}=3000$ GeV.  In Fig.~\ref{scpv_mHmax_bm1}, the
maximal values of $m_{H_i}$ ($i=1-4$) (left figure) and $|\sin\varphi_1|$ and
$|\sin\varphi_2|$ (right figure) are plotted as functions of $m_{H^\pm}$.  For
each fixed $m_{H^\pm}$, all $m^{\rm max}_{H}$ are obtained for different sets of
$(m^{}_{SS_1}, m^{}_{SS_2}, m^{}_{S_1S_2})$.  One can see that the upper bounds
on $m_{H_i}$ strongly depend on $m_{H^\pm}$ except for $m_{H_2}$.  It is found
that the upper bound on the lightest neutral Higgs boson mass $m^{\rm
  max}_{H_1}$ is below 125 GeV and can reach up to around 123 GeV for
$m_{H^\pm}=334$ GeV.  Since the lightest state $H_1$ is mainly composed of the
singlet states, $m_{H_1}$ do not increase even if we change the values of
($m_{\tilde{q}}, m_{\tilde{t}_R}, m_{\tilde{b}_R}$)=(1000, 500, 500) GeV into,
say (3000, 1500, 1500) GeV.  In this case, the second lightest Higgs boson $H_2$
receives corrections from the top/stop loops.  From the right figure of
Fig.~\ref{scpv_mHmax_bm1}, one can see that the $CP$ symmetry is maximally
violated when $m^{\rm max}_{H_1}>100$ GeV.

It is noticed that the $CP$-violating solutions $\varphi_1$ and $\varphi_2$ are
obtained by solving the necessary conditions for SCPV, Eqs.~(\ref{cond_scpv1})
and (\ref{cond_scpv2}).  In order to check whether they give $CP$ violation at
the vacuum, we perform the minimization in the ten-dimensional parameter space
($v_d$, $v_u$, $v_S$, $v_{S_1}$, $v_{S_2}$, $v_{S_3}$, $\theta_2$, $\theta_S$,
$\theta_{S_1}$, $\theta_{S_2}$), and find that the solutions obtained above
indeed give the $CP$-violating vacuum.

\subsection{EDM constraints}\label{subsec:EDM}
The $CP$-violating phases can also be constrained by the upper bounds on
electric dipole moments (EDMs) of electron, neutron and mercury
atom~\cite{EDM1,EDM2}.  Similar to the MSSM, the SUSY particles-mediated
one-loop diagrams contribute to the EDMs.  However, we assume that the only
sources of $CP$ violation come from $\theta_{S_1S_2}$ for ECPV and
$\varphi_i~(i=1,2)$ for SCPV in the sMSSM.  Therefore, their contributions to
the EDMs generally vanish.  At the two-loop level, however, the Higgs bosons
with indefinite $CP$ properties can contribute to the so-called Barr-Zee type
diagrams~\cite{EDM2} and become sizable when $\tan\beta$ is large.  Since we
take $\tan\beta=1$ in the $CP$-violating cases, we expect that they do not put
severe constraints on $\theta_{S_1S_2}$ or $\varphi_i~(i=1,2)$.

%
%
\section{Conclusions}\label{sec:conclusion}
We have studied the Higgs sector of the sMSSM with particular focus on $CP$
violation.  The masses and couplings of the Higgs bosons are calculated using
the one-loop effective potential, including corrections due to the
third-generation quarks and squarks.  Imposing both the theoretical and
experimental constraints, the allowed region is obtained for Case I and Case II
defined in the text.  In short, all Higgs VEVs of the secluded Higgs singlets in
Case I are taken to be of $\mathcal{O}$(TeV), and in Case II two of them are of
$\mathcal{O}$(100 GeV) and the other of $\mathcal{O}$(TeV).  Due to the
corrections from the Higgs singlets, the $\tan\beta=1$ case cannot be ruled out
by the LEP experimental results.  However, the conditions for the desired
electroweak vacuum generally render a very restrictive parameter space.

In this model, ECPV can be induced by the nonzero phase of $m^2_{S_1S_2}$ at the
tree level.  It is found that a large value of $\theta_{S_1S_2}$ can make the
lightest Higgs boson lighter than the LEP bound of 114.4 GeV, provided that the
Higgs coupling to the $Z$ boson is sufficiently suppressed, similar to the CPX
scenario in the MSSM.  Nevertheless, large $\mu$ and $A$ terms are not required
in the sMSSM for the realization of large $CP$ violation.  Therefore, the
spectrum of SUSY particles is generally different from the MSSM CPX scenario.

We have also investigated the SCPV scenario.  Unlike the MSSM, SCPV can occur at
the tree level in the presence of the nonzero $\theta$'s residing in the
quadratic terms of the Higgs potential.  Our analysis shows that in this case
the lightest Higgs boson mass has a certain upper bound, depending on the
charged Higgs boson mass.  In a specific case, the maximal value of $m_{H_1}$ is
around 125 GeV for $m_{H^\pm}=334$ GeV with the $CP$-violating phases being
nearly maximal.

In this paper, it is assumed that the only sources of $CP$ violation come from
the Higgs sector.  Such $CP$-violating phases show up in the Higgs
boson-mediated two-loop diagrams that contribute to the EDMs of electron,
neutron and mercury atom.  However, these diagrams are not important as long as
$\tan\beta=1$.

As pointed out in Ref.~\cite{EWBG_sMSSM}, a strong first order electroweak phase
transition is possible in the sMSSM due to the presence of the trilinear term
$\lambda A_\lambda S\Phi_d\Phi_u$.  In this case, the light stop is not
necessarily lighter than the top quark as required in the MSSM.  A devoted study
of the electroweak phase transition with/without $CP$ violation will be
presented elsewhere~\cite{CS}.

%
%
\appendix
%
%
\section{The mass matrix of the neutral Higgs bosons at the tree level}
\label{app:neu_MassMat}
%
%
Here we present explicitly the tree-level squared mass matrix elements for the
neutral Higgs bosons.  The $CP$-even part is given by
\begin{eqnarray}
\frac{1}{2}
\left(
\begin{array}{cc}
\bec{h}^T_O & \bec{h}^T_S
\end{array}
\right) {\cal M}_S^2
\left(
\begin{array}{c}
\bec{h}_O \\
\bec{h}_S
\end{array}
\right),\quad
{\cal M}_S^2=
\left(
\begin{array}{cc}
{\cal M}_S^{(O)} & {\cal M}_{S}^{(OS)} \\
\Big({\cal M}_{S}^{(OS)}\Big)^T & {\cal M}_S^{(S)}
\end{array}
\right),
\end{eqnarray}
where
\begin{eqnarray}
({\cal M}_S^{(O)})_{11}&=&\left[\frac{g_2^2+g_1^2}{4}+g'^2_1Q_{H_d}^2\right]v_d^2
	+R_\lambda\frac{v_uv_S}{v_d}, \\
({\cal M}_S^{(O)})_{22}&=&\left[\frac{g_2^2+g_1^2}{4}+g'^2_1Q_{H_u}^2\right]v_u^2
	+R_\lambda\frac{v_dv_S}{v_u}, \\
({\cal M}_S^{(O)})_{33}&=&{\rm Re}(m_{SS_1}^2e^{i\varphi_1})\frac{v_{S_1}}{v_S}
	+{\rm Re}(m_{SS_2}^2e^{i\varphi_2})\frac{v_{S_2}}{v_S}+R_\lambda\frac{v_dv_u}{v_S}
	+g'^2_1Q_S^2v_S^2, \\
({\cal M}_S^{(O)})_{12}&=&({\cal M}_S^{(O)})_{21}
	=\left[-\frac{g_2^2+g_1^2}{4}+|\lambda|^2+g'^2_1Q_{H_d}Q_{H_u}\right]v_dv_u
	-R_\lambda v_S, \\
({\cal M}_S^{(O)})_{13}&=&({\cal M}_S^{(O)})_{31}
	=-R_\lambda v_u+(|\lambda|^2+g'^2_1Q_{H_d}Q_S)v_dv_S,\\
({\cal M}_S^{(O)})_{23}&=&({\cal M}_S^{(O)})_{32}
	=-R_\lambda v_d+(|\lambda|^2+g'^2_1Q_{H_u}Q_S)v_uv_S,\\
({\cal M}_S^{(S)})_{11}&=&{\rm Re}(m_{SS_1}^2e^{i\varphi_1})\frac{v_{S}}{v_{S_1}}
	+{\rm Re}(m_{S_1S_2}^2e^{i\varphi_{12}})\frac{v_{S_2}}{v_{S_1}}
	+R_{\lambda_S}\frac{v_{S_2}v_{S_3}}{v_{S_1}}+g'^2_1Q_{S_1}^2v_{S_1}^2, \\
({\cal M}_S^{(S)})_{22}&=&{\rm Re}(m_{SS_2}^2e^{i\varphi_2})\frac{v_{S}}{v_{S_2}}
	+{\rm Re}(m_{S_1S_2}^2e^{i\varphi_{12}})\frac{v_{S_1}}{v_{S_2}}
	+R_{\lambda_S}\frac{v_{S_1}v_{S_3}}{v_{S_2}}+g'^2_1Q_{S_2}^2v_{S_2}^2, \\
({\cal M}_S^{(S)})_{33}&=&R_{\lambda_S}\frac{v_{S_1}v_{S_2}}{v_{S_3}}
	+g'^2_1Q_{S_3}^2v_{S_3}^2, \\
({\cal M}_S^{(S)})_{12}&=&({\cal M}_S^{(S)})_{21}
	=-{\rm Re}(m_{S_1S_2}^2e^{i\varphi_{12}})-R_{\lambda_S}v_{S_3}
	+(|\lambda_S|^2+g'^2_1Q_{S_1}Q_{S_2})v_{S_1}v_{S_2},\non\\\\
({\cal M}_S^{(S)})_{13}&=&({\cal M}_S^{(S)})_{31}
	=-R_{\lambda_S}v_{S_2}+(|\lambda_S|^2+g'^2_1Q_{S_1}Q_{S_3})v_{S_1}v_{S_3},\\	
({\cal M}_S^{(S)})_{23}&=&({\cal M}_S^{(S)})_{32}
	=-R_{\lambda_S}v_{S_1}+(|\lambda_S|^2+g'^2_1Q_{S_2}Q_{S_3})v_{S_2}v_{S_3},\\	
({\cal M}_S^{(OS)})_{11}&=&g'^2_1Q_{H_d}Q_{S_1}v_dv_{S_1},\\
({\cal M}_S^{(OS)})_{22}&=&g'^2_1Q_{H_u}Q_{S_2}v_uv_{S_2},\\
({\cal M}_S^{(OS)})_{33}&=&g'^2_1Q_SQ_{S_3}v_Sv_{S_3},\\
({\cal M}_S^{(OS)})_{12}&=&g'^2_1Q_{H_d}Q_{S_2}v_dv_{S_2},\\
({\cal M}_S^{(OS)})_{13}&=&g'^2_1Q_{H_d}Q_{S_3}v_dv_{S_3},\\
({\cal M}_S^{(OS)})_{21}&=&g'^2_1Q_{H_u}Q_{S_1}v_uv_{S_1},\\
({\cal M}_S^{(OS)})_{23}&=&g'^2_1Q_{H_u}Q_{S_3}v_uv_{S_3},\\
({\cal M}_S^{(OS)})_{31}&=&-{\rm Re}(m_{SS_1}^2e^{i\varphi_1})+g'^2_1Q_SQ_{S_1}v_Sv_{S_1},\\
({\cal M}_S^{(OS)})_{32}&=&-{\rm Re}(m_{SS_2}^2e^{i\varphi_2})+g'^2_1Q_SQ_{S_2}v_Sv_{S_2}.
\end{eqnarray}
%
%
The $CP$-odd part is given by
\begin{eqnarray}
\frac{1}{2}
\left(
\begin{array}{cc}
\bec{a}^T_O & \bec{a}^T_S
\end{array}
\right) {\cal M}_P^2
\left(
\begin{array}{c}
\bec{a}_O \\
\bec{a}_S
\end{array}
\right),\quad
{\cal M}_P^2=
\left(
\begin{array}{cc}
{\cal M}_P^{(O)} & {\cal M}_{P}^{(OS)} \\
\Big({\cal M}_{P}^{(OS)}\Big)^T & {\cal M}_P^{(S)}
\end{array}
\right),
\end{eqnarray}
where
\begin{eqnarray}
{\cal M}_P^{(O)}&=&\frac{R_\lambda v_S}{v_dv_u}
\left(
\begin{array}{ccc}
v_u^2 & v_dv_u & \frac{v_dv^2_u}{v_S} \\
v_dv_u & v_d^2 & \frac{v^2_dv_u}{v_S} \\
\frac{v_dv^2_u}{v_S} & \frac{v^2_dv_u}{v_S} & ({\cal M}_P^{(O)})_{33}
\end{array}
\right), \quad {\cal M}_P^{(OS)}= 
\left(
\begin{array}{ccc}
0 & 0 & 0 \\
0 & 0 & 0 \\
{\rm Re}(m_{SS_1}^2e^{i\varphi_1}) & {\rm Re}(m_{SS_2}^2e^{i\varphi_2}) & 0 \\
\end{array}
\right), \non\\
{\cal M}_P^{(S)}&=&
\left(
\begin{array}{ccc}
({\cal M}_P^{(S)})_{11} &
-{\rm Re}(m_{S_1S_2}^2e^{i\varphi_{12}})+R_{\lambda_S}v_{S_3} & R_{\lambda_S}v_{S_2} \\
-{\rm Re}(m_{S_1S_2}^2e^{i\varphi_{12}})+R_{\lambda_S}v_{S_3} &
({\cal M}_P^{(S)})_{22} & R_{\lambda_S}v_{S_1} \\
R_{\lambda_S}v_{S_2} & R_{\lambda_S}v_{S_1} & R_{\lambda_S}\frac{v_{S_1}v_{S_2}}{v_{S_3}}
\end{array}
\right), 
\end{eqnarray}
with
\begin{eqnarray}
({\cal M}_P^{(O)})_{33} &=&{\rm Re}(m_{SS_1}^2e^{i\varphi_1})\frac{v_{S_1}}{v_S}
	+{\rm Re}(m_{SS_2}^2e^{i\varphi_2})\frac{v_{S_2}}{v_S}
	+R_{\lambda}\frac{v_dv_u}{v_S},\\
({\cal M}_P^{(S)})_{11} &=&{\rm Re}(m_{SS_1}^2e^{i\varphi_1})\frac{v_{S}}{v_{S_1}}
	+{\rm Re}(m_{S_1S_2}^2e^{i\varphi_{12}})\frac{v_{S_2}}{v_{S_1}}
	+R_{\lambda_S}\frac{v_{S_2}v_{S_3}}{v_{S_1}},\\
({\cal M}_P^{(S)})_{22} &=&{\rm Re}(m_{SS_2}^2e^{i\varphi_2})\frac{v_{S}}{v_{S_2}}
	+{\rm Re}(m_{S_1S_2}^2e^{i\varphi_{12}})\frac{v_{S_1}}{v_{S_2}}
	+R_{\lambda_S}\frac{v_{S_1}v_{S_3}}{v_{S_2}}.
\end{eqnarray}
%
%
The mixing between $CP$-even and $CP$-odd parts is already given in the main
text.

\begin{acknowledgments}
  We would like to thank Koichi Funakubo and C.-P. Yuan for useful discussions
  and comments.  This work is supported in part by the National Science Council
  of Taiwan, R.O.C. under Grant No.\ NSC~96-2112-M-008-001.
\end{acknowledgments}

\end{document}